\documentclass[conference]{IEEEtran}
\IEEEoverridecommandlockouts
\usepackage{cite}
\usepackage{xspace}
\usepackage{amsmath,amssymb,amsfonts}
\usepackage[breakable, skins]{tcolorbox}
\usepackage{algorithmic}
\usepackage{graphicx}
\usepackage{textcomp}
\usepackage{xcolor}
\usepackage{caption}
\usepackage{subcaption}
\usepackage{multirow}
\usepackage{colortbl}
\usepackage{booktabs}
\usepackage{color, soul}
\usepackage{hhline}
\usepackage{tabularx}
\usepackage{adjustbox}
\usepackage[hyphens]{url}
\usepackage{enumitem}
\usepackage{balance}
\usepackage{dblfloatfix}
\usepackage[absolute,showboxes]{textpos}

\tcbset{arc=0mm,size=fbox,attach title to upper={\ ---\ },coltitle=black}

\usepackage{hyperref}
\hypersetup{pdfstartview=FitV,
  breaklinks=true, colorlinks=true, linkcolor=black, citecolor=black, urlcolor=blue,
  pdftitle={From Single Lane to Highways: Analyzing the Adoption of Multipath TCP in the Internet}, pdfauthor={}, pdfkeywords={}
}

\usepackage{cleveref}
\crefformat{section}{\S#2#1#3}
\crefformat{subsection}{\S#2#1#3}
\crefformat{subsubsection}{\S#2#1#3}
\crefrangeformat{section}{\S\S#3#1#4 to~#5#2#6}
\crefmultiformat{section}{\S\S#2#1#3}{ and~#2#1#3}{, #2#1#3}{ and~#2#1#3}

\usepackage{todonotes} %
\presetkeys{todonotes}{inline,color=olive!30}{}

\newcommand{\zmap}{ZMap\xspace}
\newcommand{\tracebox}{Tracebox\xspace}

\newcommand{\eg}{e.g.,\xspace}
\newcommand{\ie}{i.e.,\xspace}
\newcommand{\etal}{\textit{et al.}\xspace}
\newcommand{\one}{(1)\xspace}
\newcommand{\two}{(2)\xspace}
\newcommand{\three}{(3)\xspace}
\newcommand{\four}{(4)\xspace}

\begin{document}
\bstctlcite{IEEEexample:BSTcontrol}

\title{From Single Lane to Highways: Analyzing\\ the Adoption of Multipath TCP in the Internet}

\author{
  \IEEEauthorblockN{Florian Aschenbrenner\IEEEauthorrefmark{1}, Tanya Shreedhar\IEEEauthorrefmark{2}, Oliver Gasser\IEEEauthorrefmark{4}, Nitinder Mohan\IEEEauthorrefmark{1}, Jörg Ott\IEEEauthorrefmark{1}}
\IEEEauthorblockA{\IEEEauthorrefmark{1}\textit{TUM} Germany\hspace{1em}\IEEEauthorrefmark{2}\textit{IIIT-Delhi} India \hspace{1em}
\IEEEauthorrefmark{4}\textit{MPI-Informatics} Germany
  }
}

\IEEEoverridecommandlockouts\IEEEpubid{\makebox[\columnwidth]{ISBN 978-3-903176-39-3~\copyright~2021 IFIP \hfill} \hspace{\columnsep}\makebox[\columnwidth]{ }}

\maketitle

\begin{abstract}

Multipath TCP (MPTCP) extends traditional TCP to enable simultaneous use of multiple connection endpoints at the source and destination. MPTCP has been under active development since its standardization in 2013, and more recently in February 2020, MPTCP was upstreamed to the Linux kernel.

In this paper, we provide the first broad analysis of MPTCPv0 in the Internet. We probe the entire IPv4 address space and an IPv6 hitlist to detect MPTCP-enabled systems operational on port 80 and 443. Our scans reveal a steady increase in MPTCP-capable IPs, reaching 9k+ on IPv4 and a few dozen on IPv6. We also discover a significant share of seemingly MPTCP-capable hosts, an artifact of middleboxes mirroring TCP options. We conduct targeted HTTP(S) measurements towards select hosts and find that middleboxes can aggressively impact the perceived quality of applications utilizing MPTCP. Finally, we analyze two complementary traffic traces from CAIDA and MAWI to shed light on the real-world usage of MPTCP. We find that while MPTCP usage has increased by a factor of 20 over the past few years, its traffic share is still quite low.

\end{abstract}

\setlength{\TPHorizModule}{\paperwidth}
\setlength{\TPVertModule}{\paperheight}
\TPMargin{5pt}
\begin{textblock}{0.8}(0.1,0.02)
    \noindent
    \footnotesize
    If you cite this paper, please use the IFIP Networking reference:
    Florian Aschenbrenner, Tanya Shreedhar, Oliver Gasser, Nitinder Mohan, Jörg Ott. 2021.
    From Single Lane to Highways: Analyzing the Adoption of Multipath TCP in the Internet.
    In \textit{IFIP Networking Conference, June 21--24, 2021, Virtual Event, Finland.}
\end{textblock}

\section{Introduction}\label{sec:introduction}

Despite significant advances in Internet infrastructure and connectivity, TCP's connectivity model has remained largely unchanged over the last 30 years. 
Recent advances in network technologies have led to the rise of multi-homed devices, \eg smartphones, with access to more than one networking interface.
Multipath TCP (MPTCP) is an extension to TCP that allows endpoints to simultaneously utilize multiple interfaces 
for concurrent or backup data transmissions~\cite{overviewMPTCP}.
Standardized in early 2013, MPTCP has shown better resource utilization, higher aggregated throughput, and resilience to network failures in numerous research studies published over the years.

Due to the performance benefits of MPTCP over TCP, several known organizations have incorporated the protocol within their products and services. 
Apple uses MPTCP in its iOS devices to enhance user experience surrounding its system services, \eg 
Siri, Music, Maps, Wi-Fi Assist~\cite{apple_backup}.
In 2019, Apple provided APIs to third-party developers for making use of MPTCP in non-system iOS applications.
Korea Telecom, in partnership with Samsung, uses MPTCP to provide Gigabit speeds over Wi-Fi and LTE to its customers~\cite{kt-gigalte}.
In February 2020, MPTCPv1 was upstreamed to Linux and is now available to all users running Linux 5.6 or newer~\cite{upstream_linux}. 

Despite significant interest in improving the protocol~\cite{overviewMPTCP,paasch2014scheduler,shreedhar2018qaware}
, the current state of MPTCP deployment in the Internet remains largely unexplored in research.
We attribute this gap partially to the influence of middleboxes on the accuracy of such studies.
The Internet is proliferated with a wide spectrum of specialized appliances and systems known as middleboxes, that meddles with user traffic before it reaches the target~\cite{sherry2012making}. 
The intended operation of middleboxes is to offer valuable benefits, \eg firewalls drop unintentional packets and proxies improve the performance of connection setup.
However, certain middleboxes interact quite poorly with connections containing TCP header extensions.
While some may strip the packet 
of any header additions before relaying it to the next hop, others might block the connection altogether~\cite{hesmans2013tcp}.
Since MPTCP relies on TCP extensions for signaling, it is also susceptible to such middleboxes in the Internet.
MPTCP designers incorporate several mechanisms into the protocol specification that allows the protocol to fall back to regular TCP
for data transfers, if the connection is affected by middleboxes~\cite{rfc8684}.
Despite that, middleboxes continue to hinder MPTCP studies, since %
scanning tools leverage the connection establishment mechanism to interact with targets and thus remain vulnerable to side-effects of middleboxes.    
In a study from 2015, the authors wanted to analyze the deployment of MPTCP in the Internet, it later became clear that the results include false-positives due to middleboxes echoing MPTCP options for non-MPTCP hosts~\cite{early_look}.
However, despite significant measurement challenges, assessing the adoption of MPTCP in-the-wild is still pertinent since the protocol can only be employed if there is sufficient server-side support in the Internet.

This paper presents the \emph{first} broad and multi-faceted assessment of MPTCPv0. 
We study both the \textit{infrastructure}, in terms of MPTCP-capable IPv4 and IPv6 addresses, and the \textit{traffic share} at two geographically diverse vantage points.
We identify and remove middleboxes affecting MPTCP in-the-wild and investigate if they also negatively impact MPTCP application traffic.
Specifically, our contributions are as follows. 

\begin{enumerate}[label=\Roman*., nolistsep,labelwidth=!, labelindent=0pt]
    \item We regularly probe the entire IPv4 address space and an IPv6 hitlist~\cite{gasser2018clusters} for MPTCPv0 support since July 2020 using \zmap. 
Our scans target HTTP (port 80) and HTTPS (port 443) since they make up the largest traffic share in the Internet~\cite{trevisan2020five,feldmann2020lockdown}.
We find that our scans are affected mainly by middleboxes that echo TCP extensions, indicating that traditional scanning methods are still ineffective in accurately evaluating the deployment of MPTCP.
We also observe that the number of IPs reported to support MPTCP \emph{without} replayed options increased fourfold over IPv4 port 443 since 2015.  

    \item We scrutinize targets that reportedly support MPTCP in our \zmap scans for middleboxes by using \tracebox~\cite{tracebox} and make two significant discoveries. 
First, most MPTCP-capable hosts in the IPv4 are \emph{transient}; indicating experimental connotations attached to MPTCP usage in-the-wild.
Second, we identify several middleboxes that interact in a much more complicated fashion than just echoing with MPTCP packets.
Despite that, we observe a growing adoption of MPTCP reaching 7.4k/6.9k and 31/27 over port 80/443 on IPv4 and IPv6, respectively.
 
    \item We initiate parallel HTTP(S) GET requests using MPTCP and regular TCP towards IPs identified in our \zmap and \tracebox measurements.
Our results show that a majority of \emph{truly} MPTCP-capable servers are indifferent to the choice of transport protocol for connection establishment.
However, IP addresses affected by middleboxes take longer to successfully establish a connection using MPTCP vs. TCP, hinting at the potential impact of middleboxes on MPTCP's perceived quality.

    \item We analyze usage of MPTCP for data transfers over the Internet by investigating four years of inter-domain traffic collected by CAIDA and MAWI.
Our findings show that MPTCP data usage is still quite low compared to TCP (peaking at 0.4\%), primarily due to the lack of widespread MPTCP support among clients, servers, and applications.
Despite openly supporting MPTCP, Apple's traffic share over the protocol is relatively small.
However, since the past year, we observe a steadily rising popularity of MPTCP for large data transfers.
\end{enumerate}

To foster reproducibility we publish  datasets and scripts for this work \cite{mptcpDataSet}. Additionally, we continuously perform MPTCPv0 and MPTCPv1 scans and publish the results at \url{https://mptcp.io}.
\section{Background and Related Work}\label{sec:related}

\subsection{MPTCP Connection Establishment} \label{sec:mptcp_conn}

We first detail the MPTCP connection establishment procedure 
as our measurement approach utilizes it inherently.
We refer inclined readers to~\cite{overviewMPTCP,paasch2014scheduler} for more details on MPTCP machinery, features, and design choices.

\begin{figure}[t!]
    \centering
    \includegraphics[width=0.9\columnwidth]{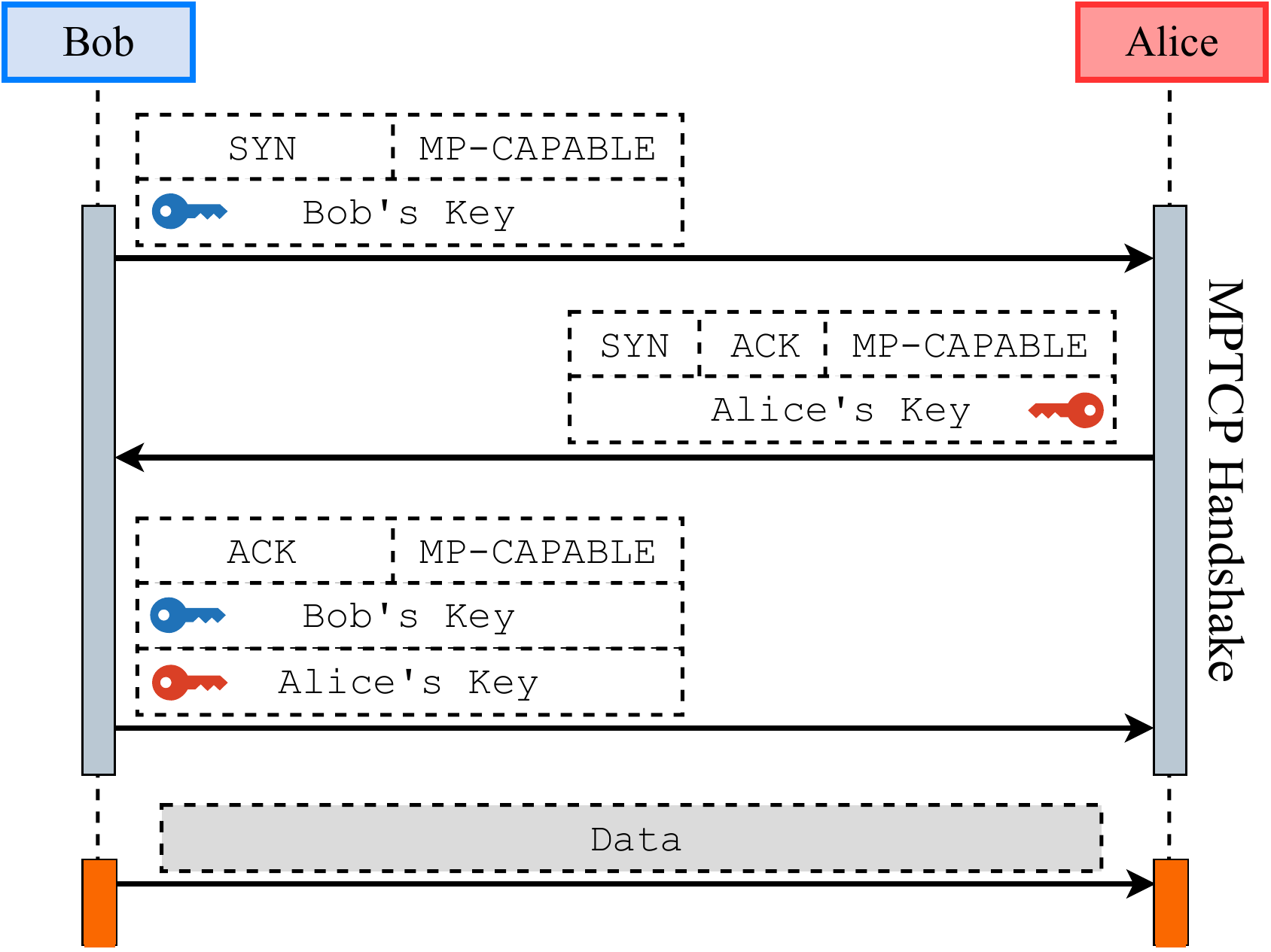}
    \caption[Initiation of MPTCP Connection]{MPTCPv0 connection establishment process between two MPTCP-capable endpoints~\cite{rfc6824}.}
    \label{fig:key_exchange}
\end{figure}

\Cref{fig:key_exchange} shows the MPTCPv0 connection establishment process between an MPTCP-enabled client and server.
The MPTCP handshake mechanism is derived from the TCP three-way handshake. 
In addition, MPTCP hosts use a random 64-bit sequence as \emph{keys} to authenticate themselves when setting up new subflows~\cite{rfc6824}. 
Moreover, every packet in the handshake signals MPTCP support through the \texttt{MP{\_}CAPABLE} option.
The client (in our example Bob) initiates a connection by sending a SYN packet containing its key and the \texttt{MP{\_}CAPABLE} option 
to the server (Alice).
If the server also supports MPTCP, it replies back with a SYN-ACK including the \texttt{MP{\_}CAPABLE}  option and its own key. 
According to the specification~\cite{rfc6824}, both key values in SYN and SYN-ACK are individually referred to as Bob's and Alice's sender's key.
In the first stage of our study, we use SYN packets to probe hosts that reply with \texttt{MP{\_}CAPABLE} option in the SYN-ACK; recording their IP address and sender's key value (for more details see \Cref{sec:zmapMethod}).

Bob finally establishes the connection by sending an ACK with both keys and the \texttt{MP{\_}CAPABLE} option.
This allows regular MPTCP data transmissions between the two parties.
Please note that the handshake procedure differs in MPTCPv1~\cite{rfc8684} where Bob does not send its key in the SYN. In this paper, we study only the deployment of MPTCPv0, and for deployment results for MPTCPv1 readers are advised to the webpage: \url{https://mptcp.io}.

\subsection{Related Work} \label{subsec:related}

Scanning the activity of different protocols in the Internet has been a long-lasting interest within the network measurement research community.
In early 2008, Heidemann \etal~\cite{heidemann2008census} systematically probed a subset of 1\% of IPv4 address space with ICMP pings.
The state of active scanning research was pushed forward significantly by \zmap \cite{durumeric2013zmap}, which allows researchers to scan the entire IPv4 address space in less than an hour.
Several works have since used the tool to investigate the deployment of different protocols and applications in the Internet, \eg liveness~\cite{bano2018scanning}, TCP initial window~\cite{conf/imc/ruth2017}, and QUIC~\cite{conf/pam/jan2018}.
Others have looked into passive data traces for a different viewpoint on deployment measurements.
Richter \etal~\cite{conf/imc/richter2016} studied the IPv4 activity as observed from within the Akamai network and show that comparative active scanning studies miss up to 40\% of the hosts that contact the CDN.
Qian \etal~\cite{conf/sigcomm/qian2009} analyzed TCP behavior 
from multiple vantage points within a large tier-1 ISP.
Wan \etal~\cite{conf/imc/wan2020} discussed several factors that can impact the quality of scanning results \eg geo-location, losses, blocking etc.
We circumvent these biases to the best of our abilities by following best practices and scanning continuously for six months.
Please refer to \Cref{sec:zmapMethod} for our detailed measurement methodology.

\begin{table*}[!t]
\centering
\setlength{\extrarowheight}{0pt}
\addtolength{\extrarowheight}{\aboverulesep}
\addtolength{\extrarowheight}{\belowrulesep}
\setlength{\aboverulesep}{0pt}
\setlength{\belowrulesep}{0pt}
\begin{adjustbox}{width=\textwidth}
\begin{tabular}{ccccccccc} 
\toprule
\multicolumn{2}{c}{ \textbf{IPv4 \zmap} } & \textbf{July}  & \textbf{August}  & \textbf{September}  & \textbf{October}  & \textbf{November}  & \textbf{December}  & {\cellcolor[rgb]{0.737,0.737,0.737}}\textbf{Consistent } \\ 
\midrule
\multirow{3}{*}{\textbf{Port 80} } & Responsive Targets & 63.9M & {\cellcolor[rgb]{1,0.596,0.596}}58.8M (-8.05\%) & 57.8M (-1.7\%) & {\cellcolor[rgb]{0.557,0.816,0.557}} 61.6M (+6.7\%) & 61.2M (-0.7\%) & 61.2M (-0.01\%) & {\cellcolor[rgb]{0.737,0.737,0.737}}- \\
 & Potential MPTCP & 179.5K & 201.6K (+12.3\%) & 197.1K (-2.2\%) & 196.1K (-0.5\%) & 205.4K (+4.8\%) & 206.3K (+0.4\%) & {\cellcolor[rgb]{0.737,0.737,0.737}}139.6K \\
 & Different Keys & 3.7K & {\cellcolor[rgb]{0.557,0.816,0.557}}4.1K (+11.1\%) & {\cellcolor[rgb]{0.557,0.816,0.557}}5K (+21.04\%) & {\cellcolor[rgb]{0.557,0.816,0.557}}8.6K (+72.7\% & 8.6K (-1.01)\% & 8.6K (+0.8)\% & {\cellcolor[rgb]{0.737,0.737,0.737}}2.6K \\ 
\hhline{---------}
\multirow{3}{*}{\textbf{Port 443} } & Responsive Targets & 47.9M & {\cellcolor[rgb]{1,0.596,0.596}}42.5M (-11.3\%) & - & {\cellcolor[rgb]{0.557,0.816,0.557}}52.9M (+24.6\%) & 52.7M (-0.3\%) & 52.6M (-0.2\%) & {\cellcolor[rgb]{0.737,0.737,0.737}}- \\
 & Potential MPTCP & 211.1K & {\cellcolor[rgb]{1,0.596,0.596}}198.1K (-6.1\%) & - & {\cellcolor[rgb]{0.557,0.816,0.557}}232.7K (+17.4\%) & 239.5K (+2.9\%) & 233.8K (-2.3\%) & {\cellcolor[rgb]{0.737,0.737,0.737}}132.4K \\
 & Different Keys & 36.6K & {\cellcolor[rgb]{1,0.596,0.596}}33.6K (-8.07\%) & - & {\cellcolor[rgb]{0.557,0.816,0.557}}46.4K (+37.9\%) & 47.1K (+1.6\%) & 47.5K (+0.9\%) & {\cellcolor[rgb]{0.737,0.737,0.737}}8.4K \\
\bottomrule
\end{tabular}
\end{adjustbox}
\caption{Summary of \zmap scans over the IPv4 address space for port 80 and port 443 from July -- December 2020. The ``Consistent'' column (colored gray) counts the common hosts that are consistently available across the measurement study.}  \label{tab:ipv4_zmap_scan}
\end{table*}

The closest work to ours dates back to 2015~\cite{early_look}.
Mehani \etal proposed a scanning mechanism that probed every host on port 80 of the Alexa Top 1M with \zmap
and classified IP addresses that responded with \texttt{MP\_CAPABLE} as supporting MPTCP.
Their results indicate that 
less than 0.1\% of scanned targets support MPTCP, with a majority located in China.
However, the accuracy of the work was later found to be low as it falsely recognized
middleboxes that echoed unknown TCP extensions as MPTCP hosts~\cite{not_easy}. The authors later published an errata and tracked the non middlebox-affected MPTCP deployment for several months in 2015.
In this work, we extend their methodology to identify middleboxes affecting MPTCP correctly, hence providing the most accurate picture of \emph{true} MPTCP deployment to date.
In addition, our study scans for MPTCP support over the two most popular services in the Internet, HTTP and HTTPS, over both IPv4 and IPv6.
\section{Active Internet Scans}\label{sec:activescans}

To identify support for Multipath TCP in the Internet, we actively scan for MPTCP options over the IPv4 and IPv6 address space.
Our study probes the entire IPv4 address space over port 80 ($
\approx$ 74M unique responsive IPs) and port 443 ($\approx$ 52M unique responsive IPs).
In IPv6, we use the \emph{IPv6 hitlist}~\cite{gasser2018clusters} to probe both port 80 and port 443 due to the size of the address space.
We find 746k and 544k responsive IPv6 addresses, respectively.
Our results are over six months of data collection from July -- December 2020.

\subsection{Methodology}\label{sec:zmapMethod}
We use \zmap~\cite{zmapv6} to rapidly enumerate IPv4 and IPv6 addresses.
To identify MPTCP hosts, we leverage the initial handshake mechanism, \ie sending a SYN with the \texttt{MP\_CAPABLE} flag along with a static sender's key.
As illustrated in \Cref{fig:key_exchange}, a legitimate MPTCP host will reply back to an MPTCP SYN with a SYN-ACK containing \texttt{MP\_CAPABLE} and its own sender's key.
If the target's SYN-ACK response includes these values, we classify it as \emph{potentially MPTCP-capable}.
Previous research has shown that the accuracy of identifying MPTCP hosts via \zmap can be low due to middleboxes that replay or strip packets with TCP extensions~\cite{early_look, not_easy, hesmans2013tcp}.
To improve the reliability of our analysis, we
probe \emph{potentially MPTCP-capable} hosts with the well-known middlebox-detection tool \tracebox~\cite{tracebox}.
\tracebox allows us to detect the presence of MPTCP options modifications on the path, revealing 
IPs that \emph{truly} support MPTCP.

Before conducting active measurements, we incorporate proposals by Partridge and Allman~\cite{partridge2016ethical} and Dittrich \etal \cite{dittrich2012menlo}.
We follow best scanning practices~\cite{durumeric2013zmap} by limiting our probing rate, maintaining a blocklist, and using dedicated servers with informing rDNS names, websites, and abuse contacts.
Furthermore, we diligently complied to any emails from organizations asking for their networks to be blacklisted.

\subsection{Finding MPTCP Support In-The-Wild} \label{sec:mptcp-zmap}

Table~\ref{tab:ipv4_zmap_scan} provides a month-wise summary of our scanning results over the IPv4 address space. 
The drop in responsive targets in August can be attributed to 
organizations that started blocking our scans---misidentifying it as a potential attack on their network. 
After taking corrective steps, the number of responsive targets increases again in October.
Unfortunately, we were unable to perform any port 443 scans in September due to infrastructural reasons.

From almost 60M responsive targets on port 80 and 50M on port 443 in IPv4, about 200k addresses responded with the \texttt{MP\_CAPABLE} flag in their SYN-ACK (\emph{potential MPTCP}), making $\approx$3.3\% and $\approx$4.6\% of total responsive hosts on port 80 and 443, respectively. 
At first glance, it might seem that MPTCP is extensively supported, more so on port 443 than on port 80.
However, we also observe that a large percentage of \emph{potential MPTCP} hosts are inconsistently active across our six-month scanning period, signaling at the existence of transient hosts.
In IPv6, we find a very small number of addresses responding with the \texttt{MP\_CAPABLE} option:
43 on TCP/80 and 165 on TCP/443.
Similar to IPv4, we see more MP\_CAPABLE addresses on port 443 (0.03\%) compared to port 80 (0.005\%).
The numbers increase to 44 and 168 until December 2020 on port 80 and 443, respectively.

\begin{figure}[t]
    \begin{subfigure}{0.5\columnwidth}
    \includegraphics[width=\textwidth]{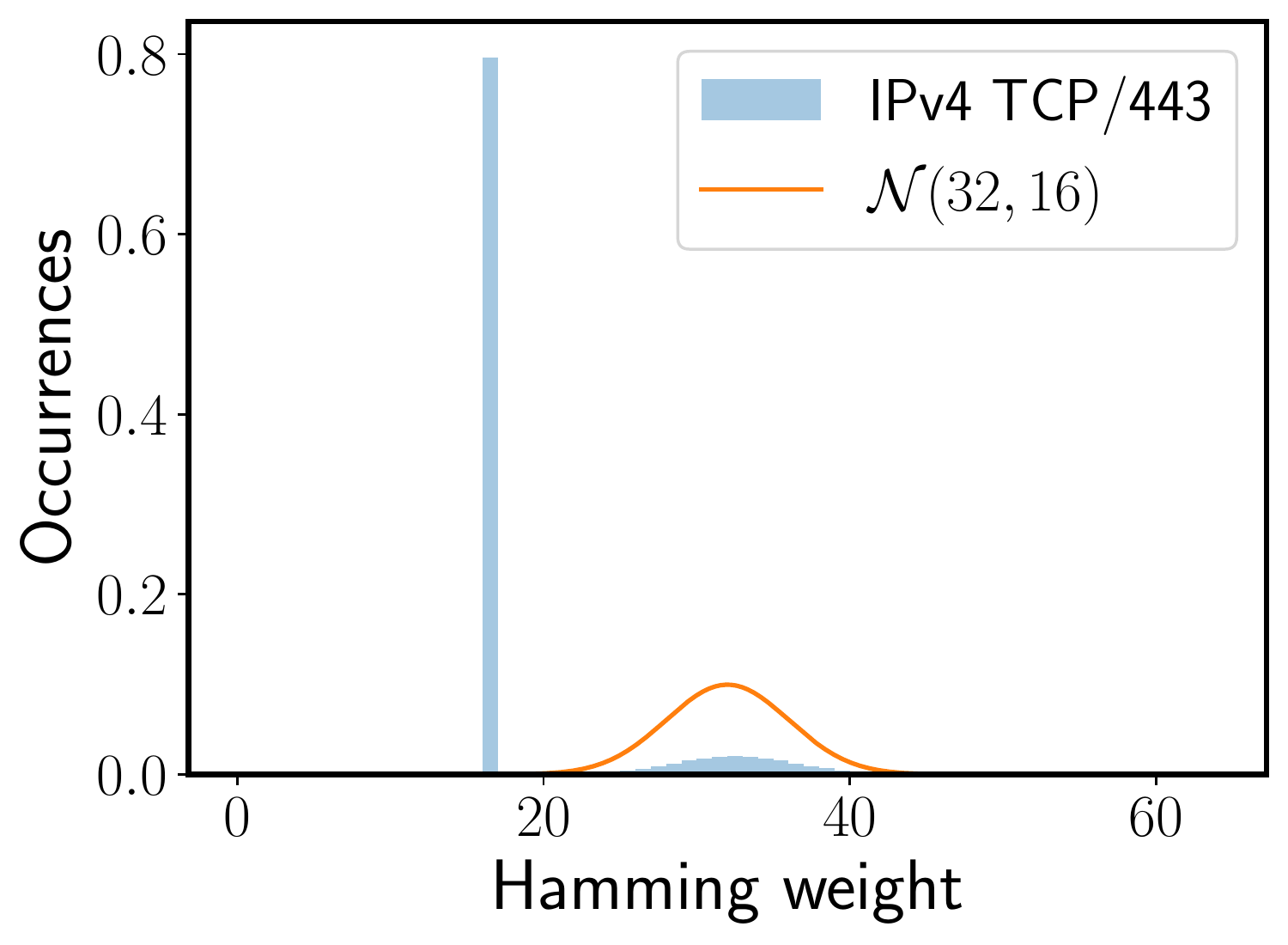}
    \caption{IPv4 Port 443}
    \label{fig:p443v4hamming}
    \end{subfigure}%
     \begin{subfigure}{0.5\columnwidth}
    \includegraphics[width=\textwidth]{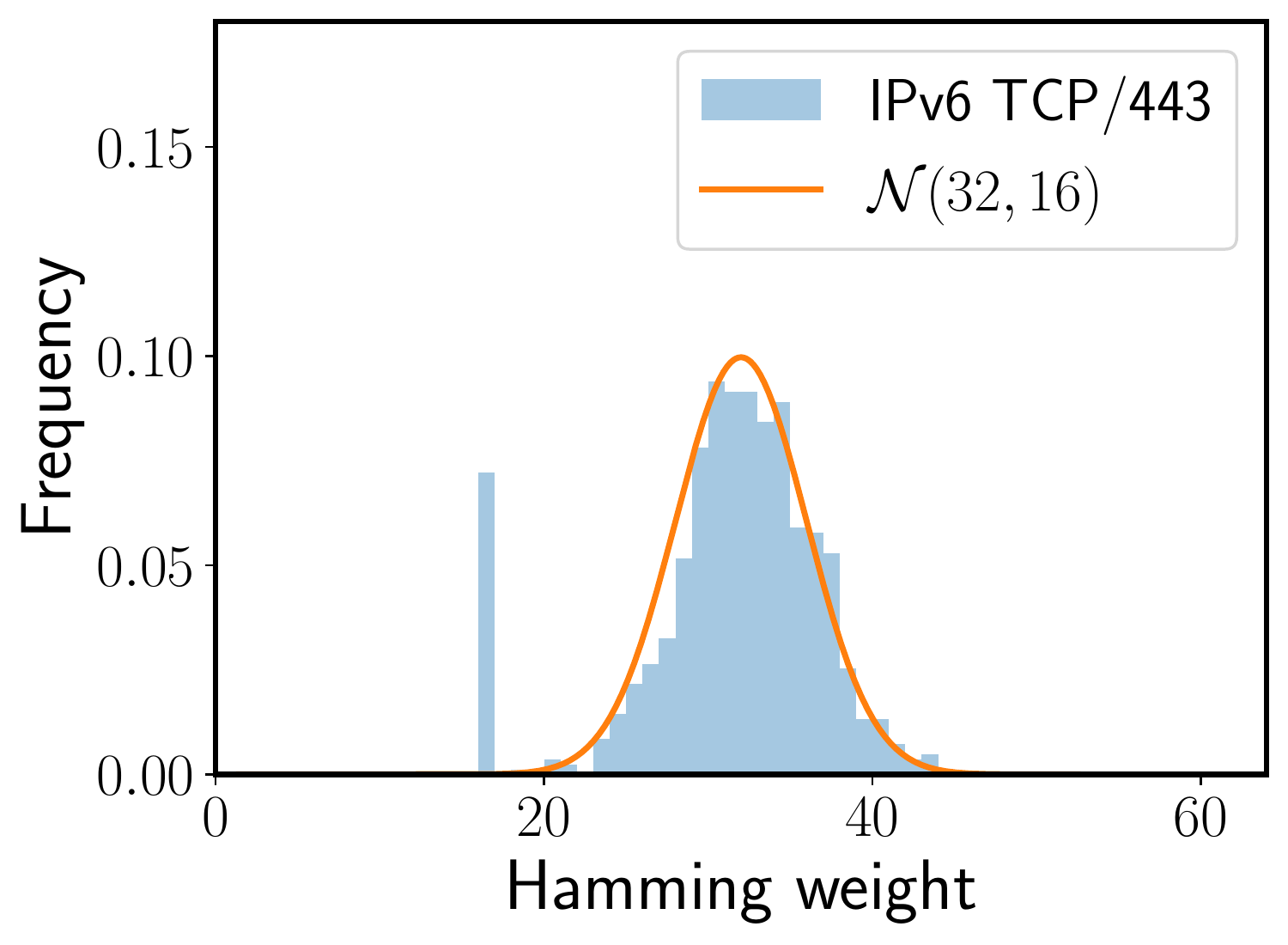}
    \caption{IPv6 Port 443}
    \label{fig:p443v6hamming}
    \end{subfigure}%
    \caption{Hamming weight distribution of sender's keys received from potential MPTCP-capable hosts in our \zmap scans.}
    \label{fig:Hamming}
\end{figure}

\smallskip
\noindent \textbf{Impact of middleboxes on correctness of scans.}
To examine if our scans are affected by interfering middleboxes, we analyze the MPTCP sender's key we receive from targets in their SYN-ACK response.
A \emph{true} MPTCP host generates a random 64-bit sequence to use as the key (see \Cref{sec:mptcp_conn}).
According to the central limit theorem, the sum of independent random variables tends toward a normal distribution.
In that case, the sum of all bits in the sender's key {---} \ie the Hamming weight {---} should follow the normal distribution $\mathcal{N}(32, 16)$.
\Cref{fig:Hamming} shows the Hamming weight distribution of the sender's key from potential MPTCP hosts on port 443 in IPv4 and IPv6.
We find that a large number of sender's keys do not follow the normal distribution.
In fact, the Hamming weight 16, \ie the exact Hamming weight of the key that we send in our SYN probes, is heavily over-represented.
This indicates a prevalence of middleboxes that mirror MPTCP options in our \zmap scans.
On port 443, the phenomenon is much more prominent in IPv4, where almost 80\% of the sender's keys are mirrored compared to $\approx$8\% of keys on IPv6.
On port 80 (not shown), we find that middlebox interference is even more elevated, with almost 90\% and 30\% of received sender's keys identified as being mirrored for IPv4 and IPv6, respectively.

\begin{figure}[t!]
\centering
    \includegraphics[width=\columnwidth]{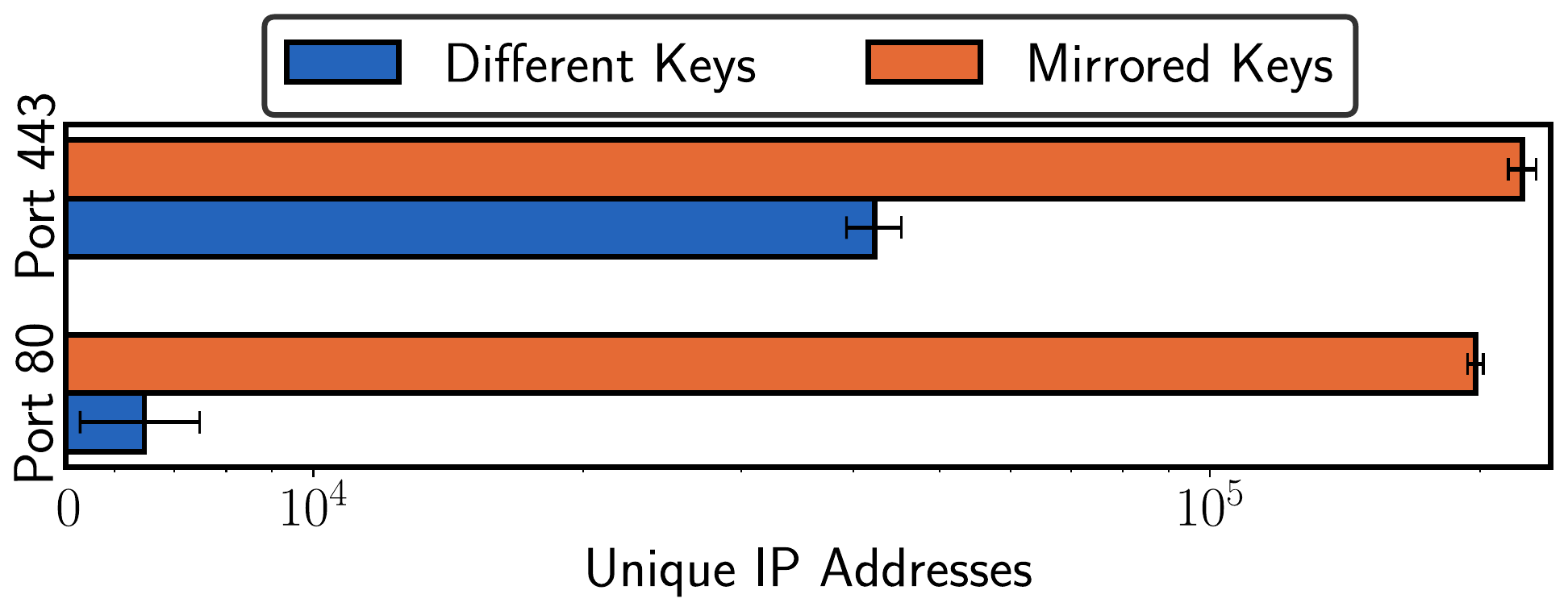}
    \caption{Unique IPv4 addresses that returned MPTCP options in our \zmap scans over port 80 and 443 categorized by the sender's key. Note that the x-axis is log-scaled.}
    \label{fig:ipv4zmap-key}
\end{figure}

We now analyze the share of hosts that are affected by middleboxes mirroring MPTCP options in our measurements.
\Cref{fig:ipv4zmap-key} shows the aggregate number of unique potential MPTCP targets scanned over IPv4 for which the sender's key was mirrored (in orange) and different from ours (in blue) for port 80 and 443.
It is evident that middleboxes affect \zmap scans quite significantly as a large percentage of hosts on both ports have mirrored keys.
Interestingly, we find that the presence of middleboxes is far greater on port 80 than on port 443, as port 80 has 96\% of hosts with mirrored keys compared to 81\% on port 443.
In contrast, 6484 and 42294 hosts send back different sender's keys on port 80 and 443, respectively.
For IPv6, we received different sender's key responses from 31 IP addresses on port 80 and 157 on port 443 (not shown).

The result is quite intriguing as it hints at HTTPS having far more support for MPTCP than HTTP over IPv4.
We also investigate whether any hosts that are middlebox-affected on port 80 are MPTCP-capable on port 443, but we find no intersection.
This leads us to believe the following contrasting possibilities.
First, HTTPS traffic is end-to-end encrypted at the application layer; it is possible that a large number of middleboxes do not modify the transport layer options of user traffic.
This results in a much smaller percentage of hosts that are affected by middleboxes that inject replayed TCP extensions.
Second, the result may still include non-MPTCP end-hosts, which are affected by middleboxes that also modify the sender's key value of SYN-ACK packets.

\begin{figure}[t!]
\centering
    \includegraphics[width=\columnwidth]{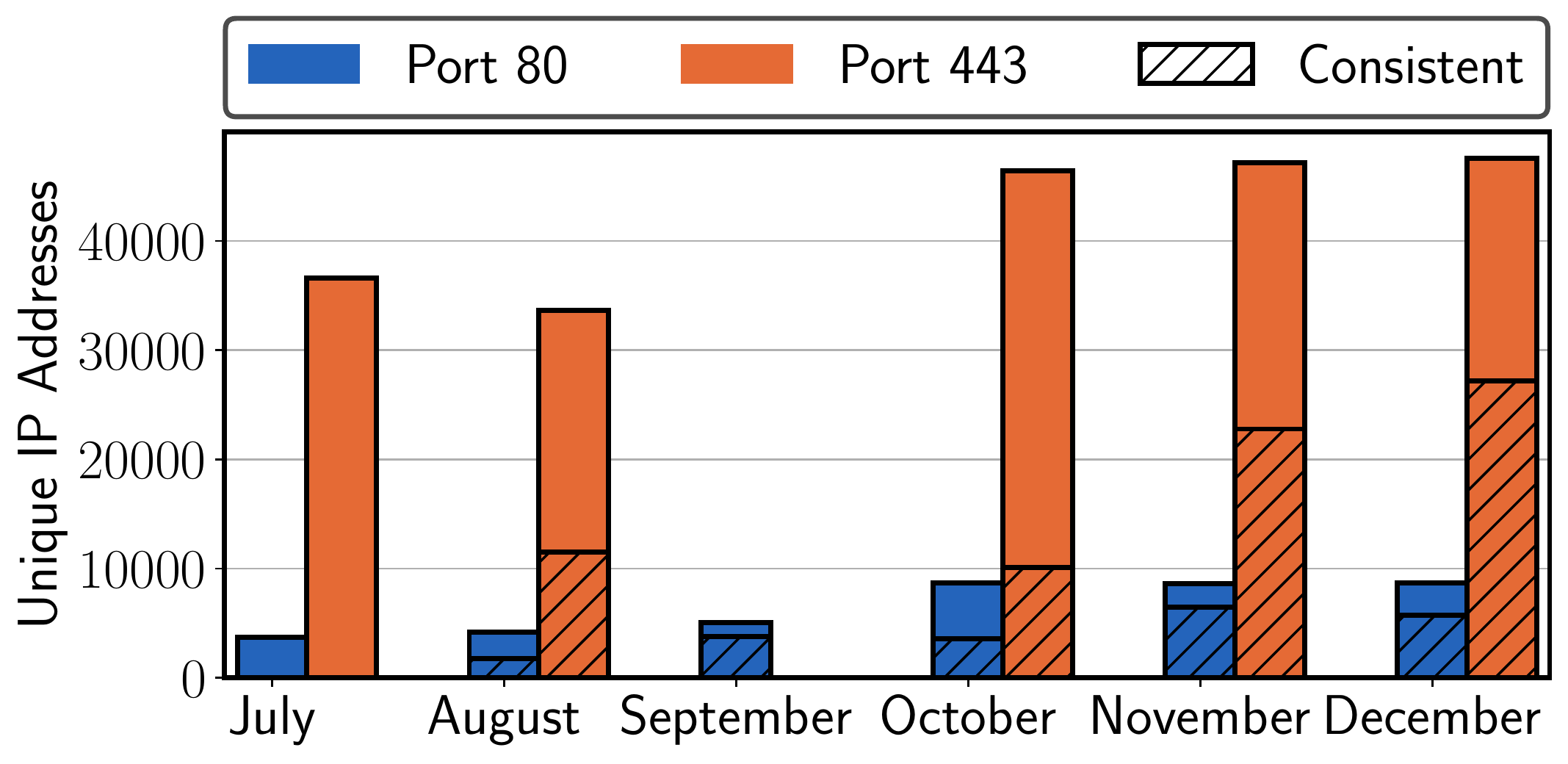}
    \caption{Monthly responsive IP addresses with different sender's keys for port 80 and 443 over IPv4 using \zmap. The \emph{shaded} region denotes IP addresses that also responded previously.}
    \label{fig:ipv4zmap-month}
\end{figure}

\begin{tcolorbox}[title=Takeaway]
    \zmap includes a large share of seemingly MPTCP-capable addresses, where middleboxes are mirroring TCP options.
    For IPs that responded with different options, we find potential MPTCP support to lean more towards HTTPS than HTTP for both IPv4 and IPv6.
\end{tcolorbox}

\subsection{Finding Interfering Middleboxes}\label{sec:tracebox}

While our ZMap analysis in the previous section filters out hosts that are affected by middleboxes simply mirroring TCP options, it still does not entirely capture the true state of MPTCP deployment in the Internet.
First, our filtering mechanism assumes that all hosts which reply with the same sender's key as ours are middlebox-affected and \emph{do not} support MPTCP.
However, this excludes the possibility of legitimate MPTCP hosts whose MPTCP options in the SYN-ACK are either stripped or overwritten by middleboxes---thus resulting in false negatives.
Second, our analysis may also include false positives due to middleboxes that may perform a complex operations on packets with extended TCP options, \eg modifying sender's keys.
Therefore, we detect the presence of interfering middleboxes by running \tracebox~\cite{tracebox} towards all targets that sent the \texttt{MP\_CAPABLE} option in our \zmap scans.%

\smallskip
\noindent \textbf{Methodology.} 
Similar to our \zmap methodology, we issue \tracebox requests with the \texttt{MP\_CAPABLE} option towards a target address.
In the reply, we receive responses from intermediate routers on the path, including any modifications made.
Overall, we observe the following different behaviors in \tracebox responses.

\begin{enumerate}[label=\Roman*., nolistsep,labelwidth=!, labelindent=0pt]
\item Only the target IP modifies the \texttt{MP\_CAPABLE} option.
\item An intermediate hop modifies the \texttt{MP\_CAPABLE} option.
\item The target was unresponsive or the query timed out.
\end{enumerate}

Based on these three categories, we classify MPTCP support as follows. 
Since 
category I responses are caused by IPs updating MPTCP options with their own sender's key in the SYN-ACK response; we classify such targets 
as \emph{truly} MPTCP-capable.
Targets in category II 
are clearly affected by middleboxes on the path and hence tagged as
\emph{middlebox-affected}.
Lastly, we classify hosts in category III as \emph{unreachable}.

\smallskip
\noindent \textbf{True MPTCP support in the Internet.}
We check if any of the end-hosts that mirror our MPTCP key in \zmap (mirrored key hosts in \Cref{fig:ipv4zmap-key}) truly support MPTCP. 
Interestingly, we did not find a single end-host that sent back the \texttt{MP\_CAPABLE} flag to our \tracebox probe for both IPv4 and IPv6. 
This confirms that our \zmap analysis does not lead to any false negatives, and checking for mirrored sender's keys is an effective first step in filtering out middleboxes.

\begin{figure}[t!]
\centering
    \includegraphics[width=\columnwidth]{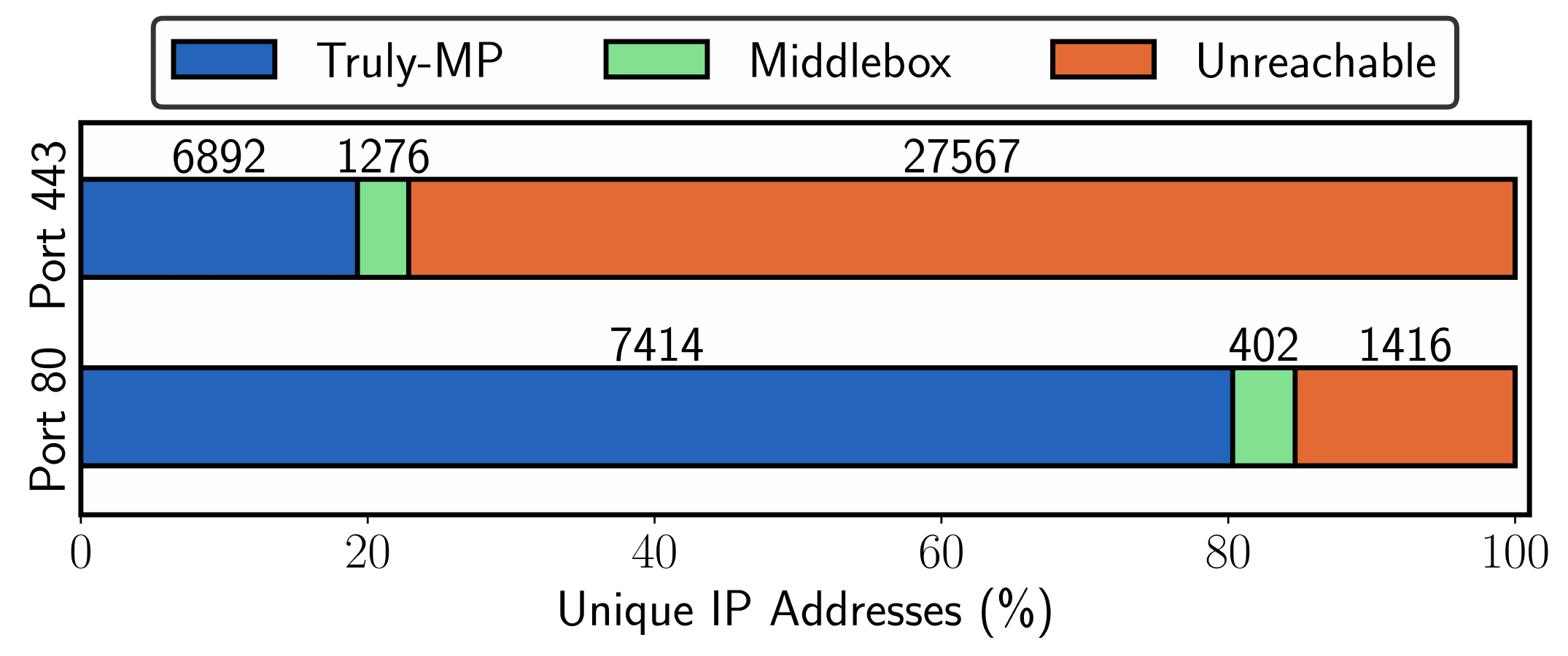}
    \caption{\tracebox analysis for \emph{consistently} responsive IPv4 addresses over port 80/443. The blue region denotes IP addresses that \emph{truly} support MPTCP, green are IP addresses affected by middleboxes on the path, and orange are unreachable.}
    \label{fig:ipv4tracebox}
\end{figure}

We continue our analysis with \tracebox responses from hosts that respond with different sender's keys in \zmap. %
We observe that a large percentage of targets do not respond to our \tracebox queries and are therefore classified as \emph{unreachable}.
This behavior is slightly more predominant in IPv4 than in IPv6, primarily due to the different target sets (Internet-wide in IPv4, hitlist-based in IPv6).
In IPv6, we only see unreachable hosts on port 443 ($\approx$ 82\%), where the majority of targets are located in the same prefix of a Dutch ISP.
These IPv6 targets respond to our \tracebox queries with \textit{Destination Unreachable (administratively prohibited)}, which hints at blocking of our queries by the ISP.
In IPv4, the number of \emph{unreachable} targets is significantly higher on port 443 ($\approx$ 90\%) than on port 80 ($\approx$~48\%).
As a result, we use \emph{consistently reachable} hosts from \zmap scans (corresponding to shaded regions in \Cref{fig:ipv4zmap-month}) for the remaining \tracebox analysis.
This precaution deflates the number of \emph{unreachable} nodes from further analysis as it removes transient hosts that are only active for short periods of time.
We confirm that the number of \emph{truly} MPTCP-capable addresses remains comparable before and after pruning transient IPs.
\Cref{fig:ipv4tracebox} shows the overview of our IPv4 \tracebox analysis for IPv4.

Despite reducing the input dataset, we observe that a large share of IPv4 targets over port 443 are still \emph{unreachable} and the absolute number of responsive addresses is similar on both port 80 and 443.
Contrary to our initial assessment based on the \zmap results, we find \emph{true} MPTCP support to be slightly higher on port 80 (7.5k hosts) than on port 443 (6.9k hosts).
Furthermore, the end-to-end encrypted nature of HTTPS does not seem to prevent middleboxes to interfere with traffic on the path as the number of \emph{middlebox-affected} hosts is $\approx$ 3$\times$ larger on port 443 compared to port 80.
In IPv6, we find not a single middlebox-affected target address.
In fact, after removing the large share of unresponsive port 443 addresses, the number of \emph{true} MPTCP-capable targets is similar on both protocols: 31 on TCP/80 and 27 on TCP/443.

\begin{figure}[t!]
\centering
    \includegraphics[width=\columnwidth]{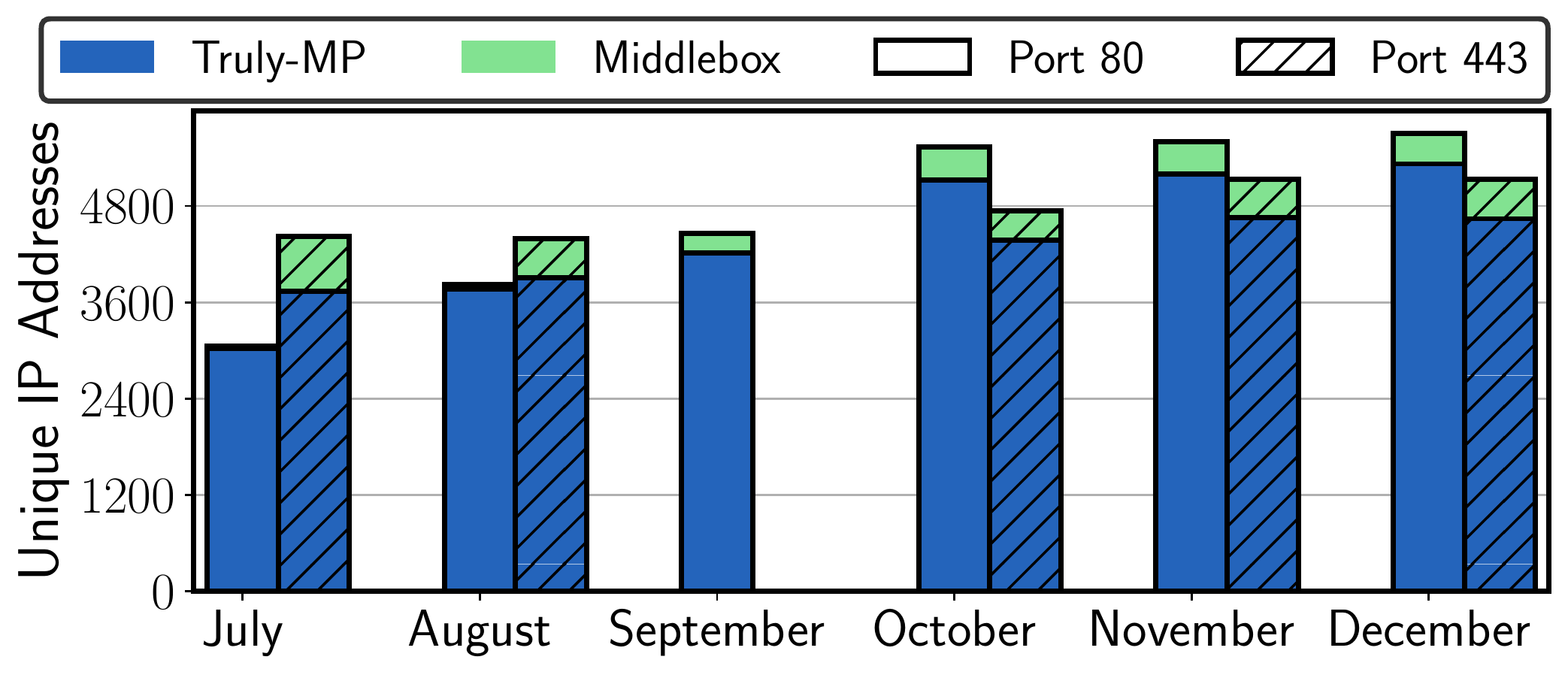}
    \caption{Monthly distribution of IPv4 addresses over port 80 and 443 that responded to our \tracebox queries.}
    \label{fig:ipv4tracebox-month}
\end{figure}

\Cref{fig:ipv4tracebox-month} shows the monthly distribution of \emph{truly} MPTCP-capable IPv4 addresses.
In IPv6, the number of MPTCP-capable addresses remains almost constant during the study period, varying only by a maximum of two addresses.
We observe that the support for MPTCP on IPv4 has been steadily increasing for both port 80 and 443 and almost doubled for port 80 over our six months study period.
As of December 2020, we identify almost $\approx$ 5.5k and $\approx$ 4.5k active MPTCP-capable IPv4 addresses on port 80 and 443, respectively.
Compared to 7.5k and 6.8k \emph{true} MPTCP hosts reported across six months (see \Cref{fig:ipv4tracebox}), the monthly distribution suggests significant transience in the MPTCP deployment.
We likely attribute these short-lived hosts to enthusiasts, system tinkerers, or researchers that may use MPTCP for short time periods.
The monthly \tracebox results are also in stark contrast to \zmap results discussed earlier (see \Cref{fig:ipv4zmap-month}) as the MPTCP support over port 80 exceeds port 443 in October 2020.

We also observe that the number of \emph{middlebox-affected} hosts remains almost consistent across months, hinting at an unvarying set of IPv4 addresses affected by interfering middleboxes on the path.
Of the 402 and 1.27k \emph{middlebox-affected} end-hosts on port 80 and 443, only 6 are found to truly support MPTCP.
However, since MPTCP options are stripped for a large fraction of \emph{middlebox-affected} end-hosts, we cannot accurately assess the true support for MPTCP within this group.
We attempt to investigate the deployment nature of middleboxes that impact MPTCP traffic using Nmap~\cite{nmap} fingerprinting.
Unfortunately, this did not lead to fruitful results due to the following hindrances.
First, the majority of middleboxes that impact our study do not respond to \tracebox probes, and hence we are unable to identify their IP address.
Second, for the handful of middleboxes that we positively identified, the accuracy of Nmap is too low (around 85\%) to confidently identify their hardware and OS characteristics. 
While most middleboxes are geo-located in China (the rest being located in Asia, North America, and Europe), their AS information does not reveal any defining traits or commonalities between organizations managing them.
We leave the thorough analysis of such middleboxes to future work.

\begin{figure}[t!]
\centering
    \includegraphics[width=\columnwidth]{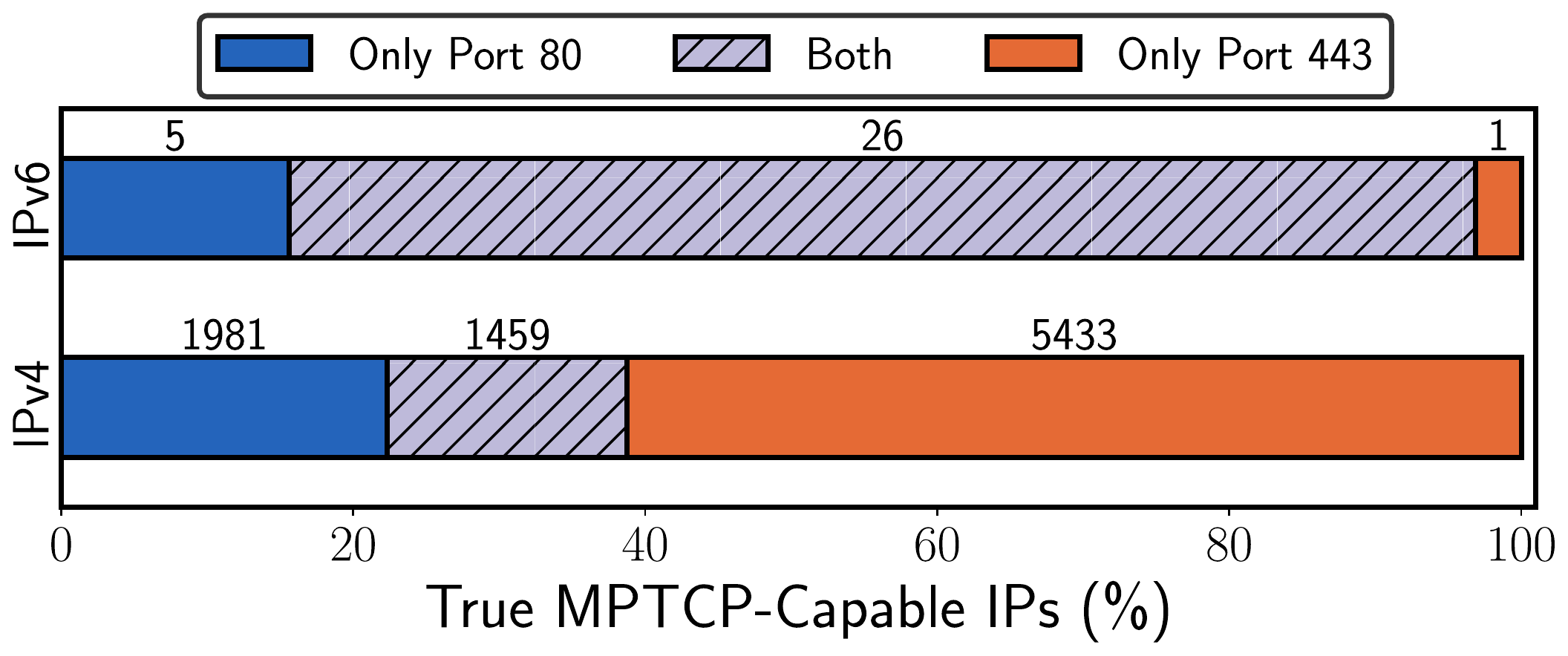}
    \caption{MPTCP support for HTTP and HTTPS in IPv4/v6.}
    \label{fig:ipv4tracebox-common}
\end{figure}

Since an MPTCP-capable machine can offer different services concurrently, we now examine the overlap between TCP/80 and TCP/443 end-hosts.
\Cref{fig:ipv4tracebox-common} shows the port breakup of all \emph{truly} MPTCP-capable IP addresses throughout our study, including transient IP addresses.
As shown in the figure, most IPv4 MPTCP hosts provide complementary services over either of the ports.
Only 16.4\% of IPv4 addresses support MPTCP over both port 80 and 443, while 22.3\% only support MPTCP over HTTP and 61\% over HTTPS.
The picture is very different for IPv6, where more than 80\% of addresses support MPTCP on both ports.

\begin{figure}[t!]
\centering
    \includegraphics[width=\columnwidth]{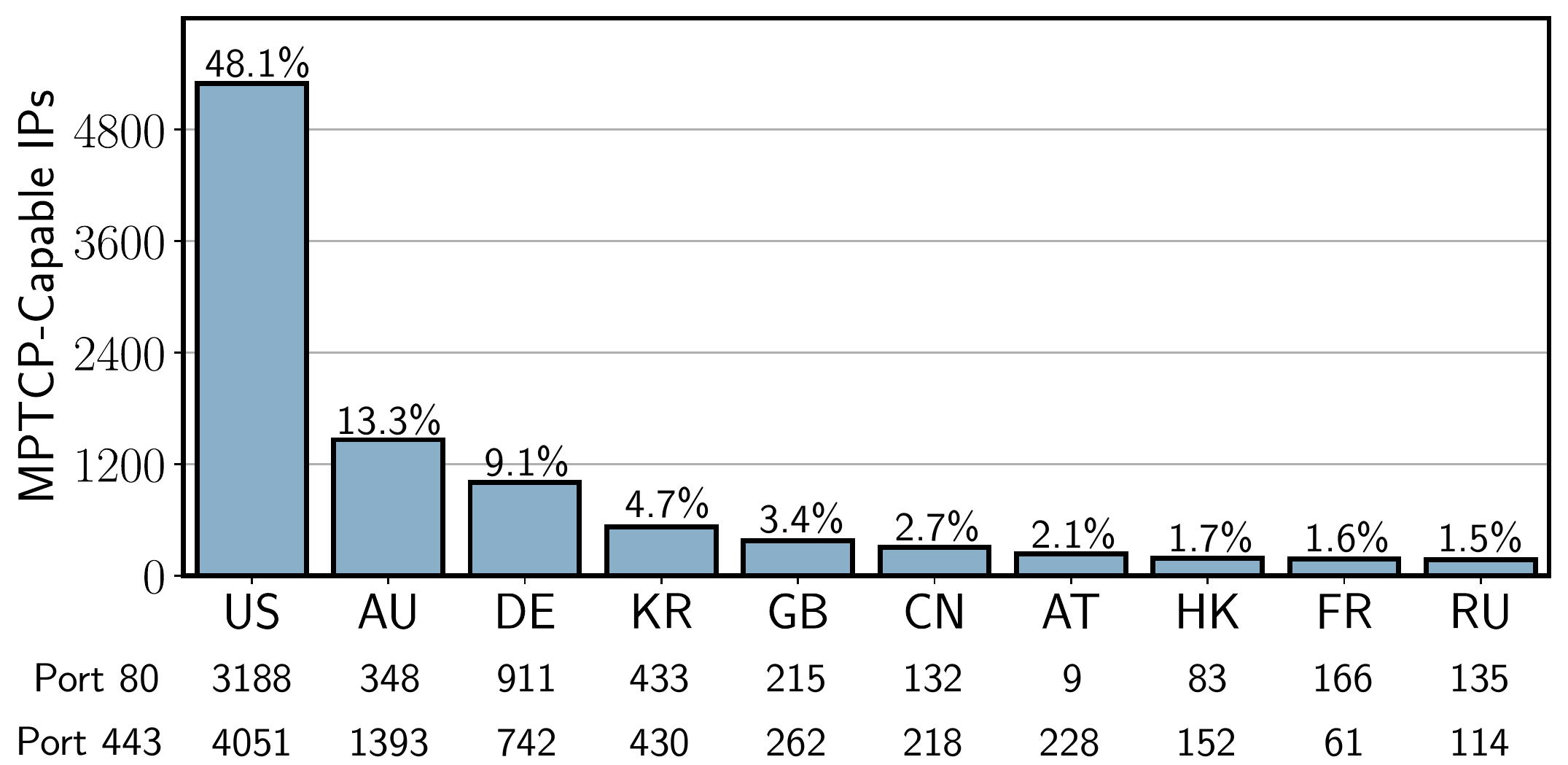}
    \caption{Geographic distribution of \emph{truly} MPTCP-capable IPv4 addresses verified by \tracebox. The bars show counts of unique IPs over both port 80 and 443 (including common IPs). The numbers below the x-axis denote the MPTCP-capable IPs serving over port 80 and 443 in that country.}
    \label{fig:ipv4tracebox-country}
\end{figure}

\smallskip
\noindent \textbf{Geo-distribution of MPTCP-capable hosts.}
We now shed some light on the physical deployment locations and operational zones of end-hosts that \emph{truly} support MPTCP.
\Cref{fig:ipv4tracebox-country} visualizes the top 10 countries with the most MPTCP-capable host densities on IPv4, arranged in decreasing fashion. 
We use the MaxMind database~\cite{maxmind} for our analysis
and only show country-level breakup since more fine-granular IP geo-location is not very accurate \cite{livadariu2020accuracy,scheitle2017hloc,shavitt2011geolocation}.
\Cref{tbl:asn-ipv4-tracebox} provides further insights, as it shows the top 10 ASes with most IPv4 MPTCP hosts on port 80 and 443.
The table also lists the associated organization name, country and AS rank, which we obtain from CAIDA's AS database~\cite{as_ranks_site}.  

\begin{table}[!t]
\setlength{\extrarowheight}{0pt}
\addtolength{\extrarowheight}{\aboverulesep}
\setlength{\aboverulesep}{0pt}
\setlength{\belowrulesep}{0pt}
\begin{adjustbox}{width=\columnwidth}
\centering
\begin{tabularx}{\columnwidth}{llllll} 
\toprule
 \textbf{ASN}~ & \textbf{\#Port80}  & \textbf{\#Port443} & \textbf{Rank}  & \textbf{Country}  & \textbf{Owner}  \\ 
\midrule
6185 & 30 & 1457 & 13577 & US & Apple Inc. \\
\rowcolor[rgb]{0.753,0.753,0.753} 1221 & 255 & 1286 & 76 & AU & Telstra Corp. \\
61157 & 674 & 499 & 1368 & DE & Plus Server \\
\rowcolor[rgb]{0.753,0.753,0.753} 7922 & 486 & 284 & 27 & US & Comcast \\
4766 & 406 & 403 & 47 & KR & Korea Telecom \\
\rowcolor[rgb]{0.753,0.753,0.753} 714 & 197 & 335 & 6630 & US & Apple Inc. \\
20115 & 239 & 205 & 98 & US & Charter Comm. \\
\rowcolor[rgb]{0.753,0.753,0.753} 18516 & 239 & 3 & 15189 & US & Molalla Comm. \\
18943 & 172 & 207 & 3855 & US & Yelcot Teleph. \\
\rowcolor[rgb]{0.753,0.753,0.753} 5607 & 116 & 116 & 5033 & GB & Sky UK Corp. \\
\bottomrule
\end{tabularx}
\end{adjustbox}
\caption{Top 10 Autonomous Systems for \emph{truly} MPTCP-enabled hosts in IPv4.}
\label{tbl:asn-ipv4-tracebox}
\end{table}

\begin{table}[!t]
\setlength{\extrarowheight}{0pt}
\addtolength{\extrarowheight}{\aboverulesep}
\setlength{\aboverulesep}{0pt}
\setlength{\belowrulesep}{0pt}
\begin{adjustbox}{width=\columnwidth}
\centering
\begin{tabularx}{\columnwidth}{llllll} 
\toprule
 \textbf{ASN}~ & \textbf{\#Port80}  & \textbf{\#Port443} & \textbf{Rank}  & \textbf{Country}  & \textbf{Owner}  \\ 
\midrule
16276 & 5 & 5 & 3623 & FR & OVH~~~~~~~~~~~ \\
\rowcolor[rgb]{0.753,0.753,0.753} 7922 & 4 & 4 & 30 & US & Comcast \\
12876 & 4 & 3 & 10887 & FR & Online S.A.S.\\
\rowcolor[rgb]{0.753,0.753,0.753} 63949 & 3 & 2 & 6706 & US & Linode \\
201155 & 2 & 2 & 25511 & CH & embeDD \\
\rowcolor[rgb]{0.753,0.753,0.753} 174 & 2 & 2 & 4  & US & Cogent \\
\bottomrule
\end{tabularx}
\end{adjustbox}
\caption{Top 6 Autonomous Systems for \emph{truly} MPTCP-enabled hosts in IPv6.}
\label{tbl:asn-ipv6-tracebox}
\end{table}

We find that almost half of all IPv4 MPTCP hosts are deployed in the US and the country dominates its closest competitors with a total of 5300 unique MPTCP-capable hosts.
\Cref{tbl:asn-ipv4-tracebox} shows that with the US, Apple has the largest deployment of MPTCP servers operational on both port 80 and 443, totaling more than 2000 unique IPv4 addresses.
The result is unsurprising since Apple has been known to publicly use MPTCP for several iOS services, \eg Siri, Music, Maps, and has recently allowed third-party developers to utilize MPTCP for non-system-native apps~\cite{apple_backup}.
The second-largest support for MPTCP over IPv4 comes from Australia, mainly due to servers hosted by Telstra, a major telecommunications company in the region.
We observe that many network operators and ISPs across the globe are utilizing MPTCP within their networks to enhance several of their client-facing services.
For example, Korea Telecom, in partnership with Samsung, uses MPTCP to provide Gigabit speeds over Wi-Fi and LTE~\cite{kt-gigalte}. 
Interestingly, we also observe from Figure~\ref{fig:ipv4tracebox-country} that in certain countries such as Austria and France, MPTCP deployment favors one port over the other, showcasing an organization's tendency to utilize MPTCP for serving specific application traffic.
In \Cref{tbl:asn-ipv6-tracebox} we show the AS distribution of \emph{truly} MPTCP-capable IPv6 addresses.
Compared to IPv4, MPTCP support in IPv6 is much more evenly distributed over ASes.
We find that most of the small number of MPTCP hosts in IPv6 are located in hosting providers and ISPs.
Overall, we find that the current MPTCP deployment spans more than 80 countries across the globe.

\begin{tcolorbox}[title=Takeaway]
   Using \tracebox we find a large share of middlebox-affected MPTCP addresses in IPv4, while IPv6 MPTCP support remains largely unaffected. Backed by Apple and major ISPs globally, IPv4 boasts of 9k+ \emph{truly} MPTCP-capable hosts, 
    compared to a few dozen on IPv6.
\end{tcolorbox}
\subsection{Middlebox Impact on MPTCP Perceived Quality} \label{subsec:mptcpqoe}

Our analysis in \Cref{sec:tracebox} revealed a widespread prevalence of middleboxes that modify extensions MPTCP relies on.  %
Previous research has shown that certain middleboxes, such as firewalls or load balancers, manipulate packets that do not fit pre-defined rule sets, \eg by marking them low-priority or forwarding them on longer paths~\cite{hesmans2013tcp}.
In this section, we want to answer whether middleboxes treat MPTCP application traffic any different from regular TCP traffic. 

We investigate this by initiating HTTP(S) GET requests using MPTCP %
from AWS in Germany towards IPv4 addresses that are marked \textit{potential}-MPTCP  in \Cref{sec:mptcp-zmap}.
We conduct the same measurements over regular TCP from the same data center in parallel.
For each successful GET response, we record \one the TCP handshake time (a.k.a. connect time), \two the TLS handshake time, \three time to first byte (TTFB), and \four the total completion time (roughly equates to website load time).
We run each measurement set, composed of 10+ runs, for almost two weeks. 
Overall, $\approx$~80\% and $\approx$~27\% targets responded to our GET requests on port 80 and 443, respectively.

 \begin{figure}[t]
	\begin{subfigure}{0.5\columnwidth}
    \includegraphics[width=\textwidth]{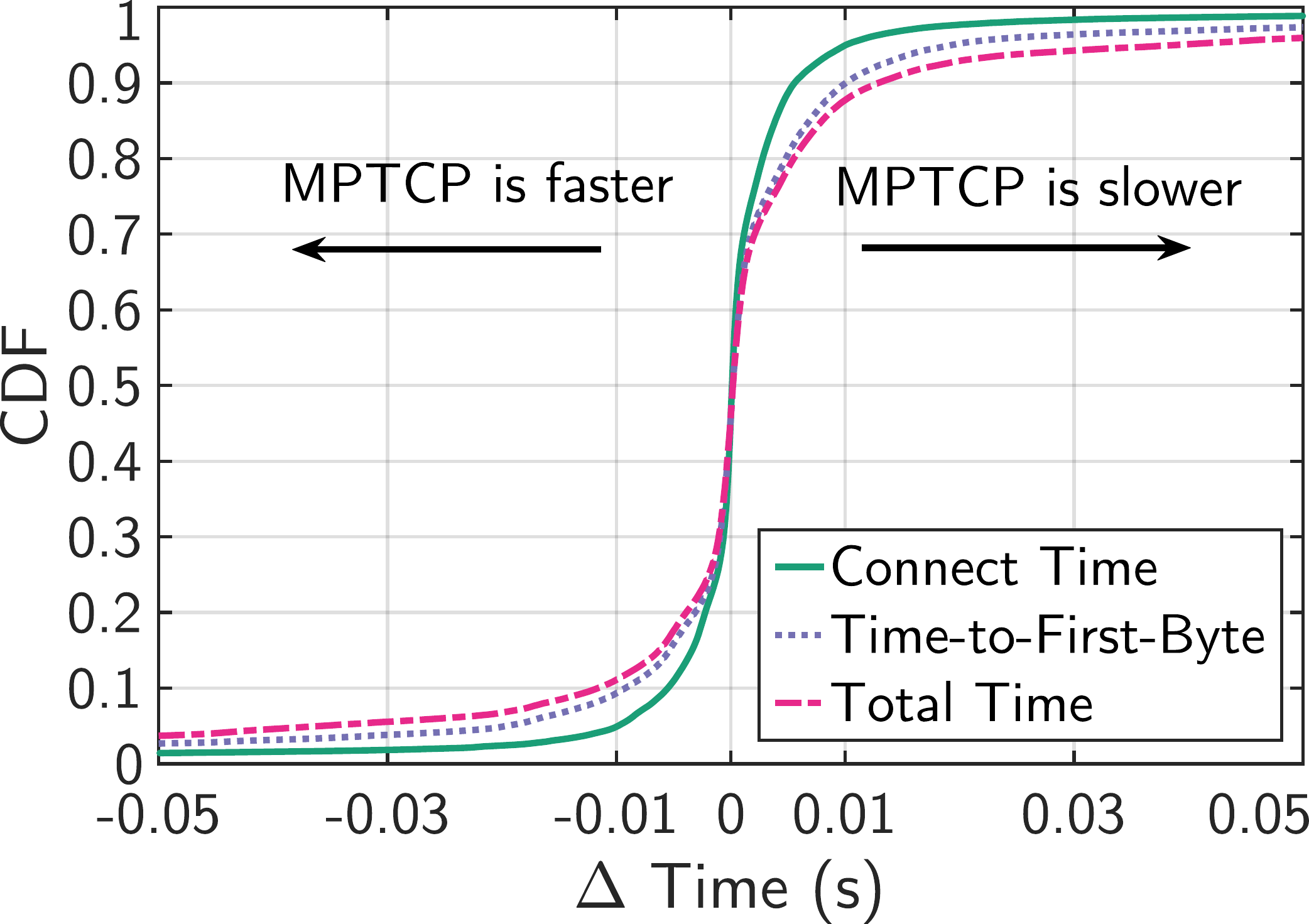}
\caption{}
    \label{fig:p80curl}
    \end{subfigure}%
    \begin{subfigure}{0.5\columnwidth}
    \includegraphics[width=\textwidth]{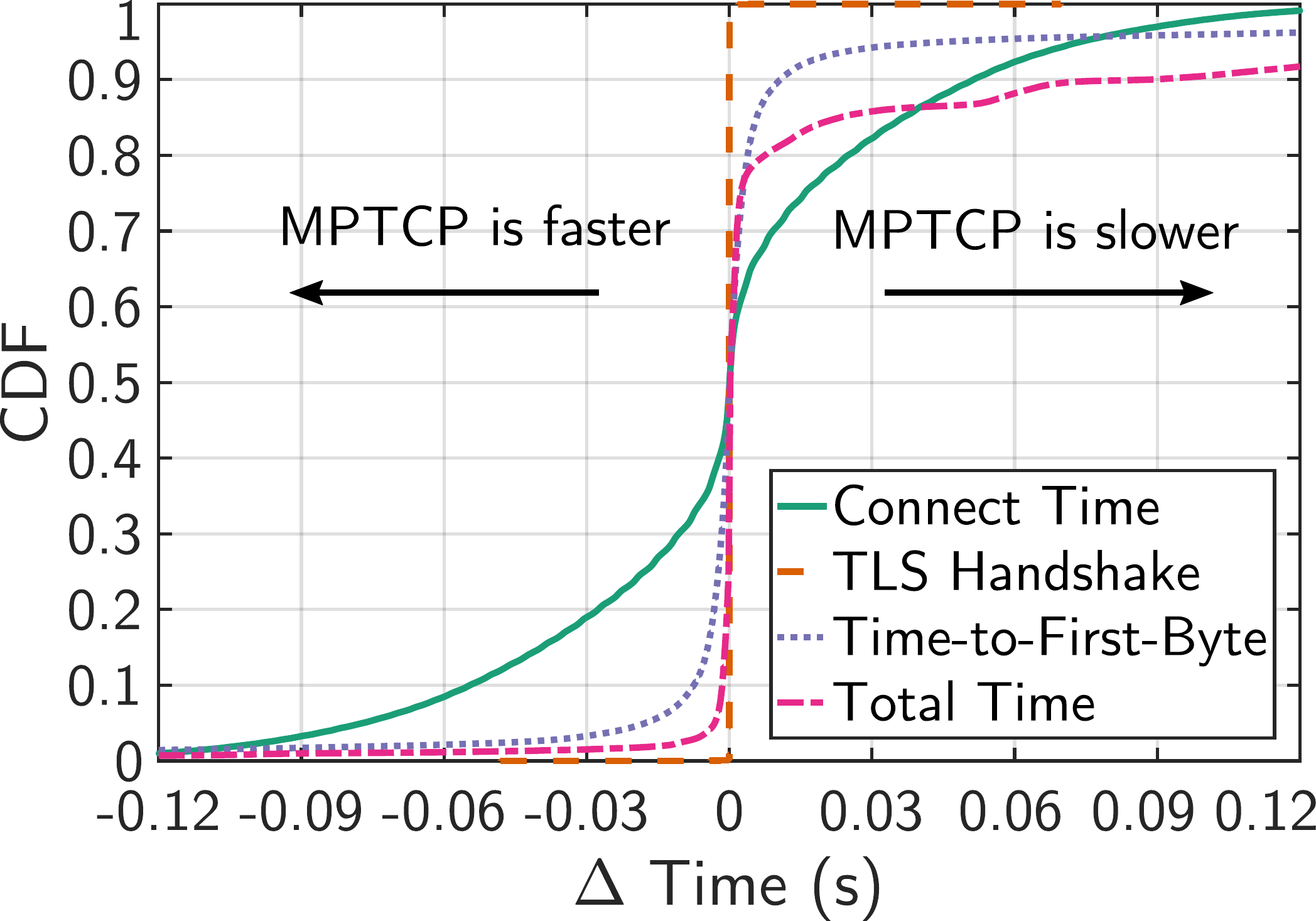}
\caption{}
    \label{fig:p443curl}
    \end{subfigure}
    \begin{subfigure}{0.5\columnwidth}
    \includegraphics[width=\textwidth]{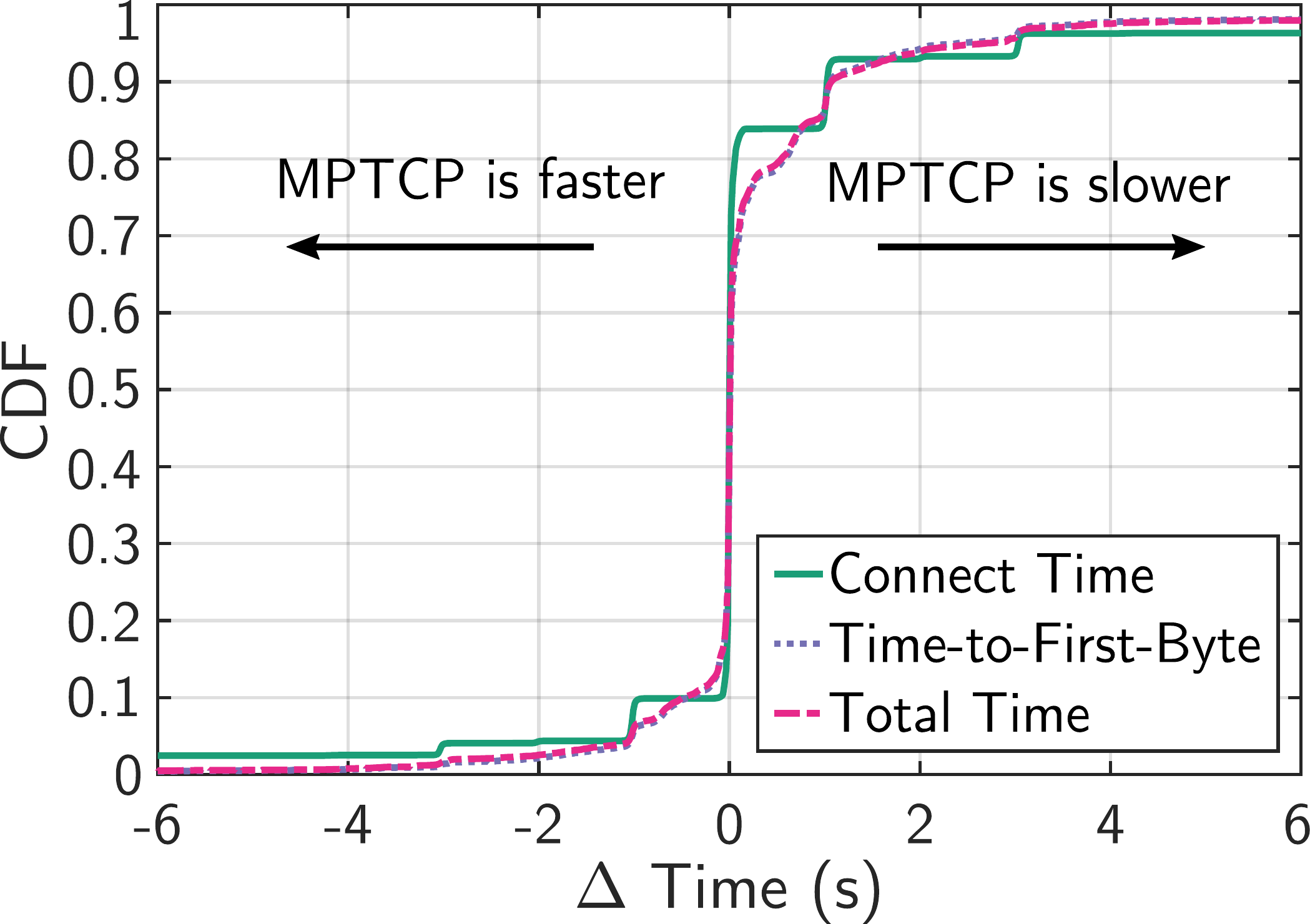}
\caption{}
    \label{fig:middleboxcurl}
    \end{subfigure}%
    	\begin{subfigure}{0.5\columnwidth}
    \includegraphics[width=\textwidth]{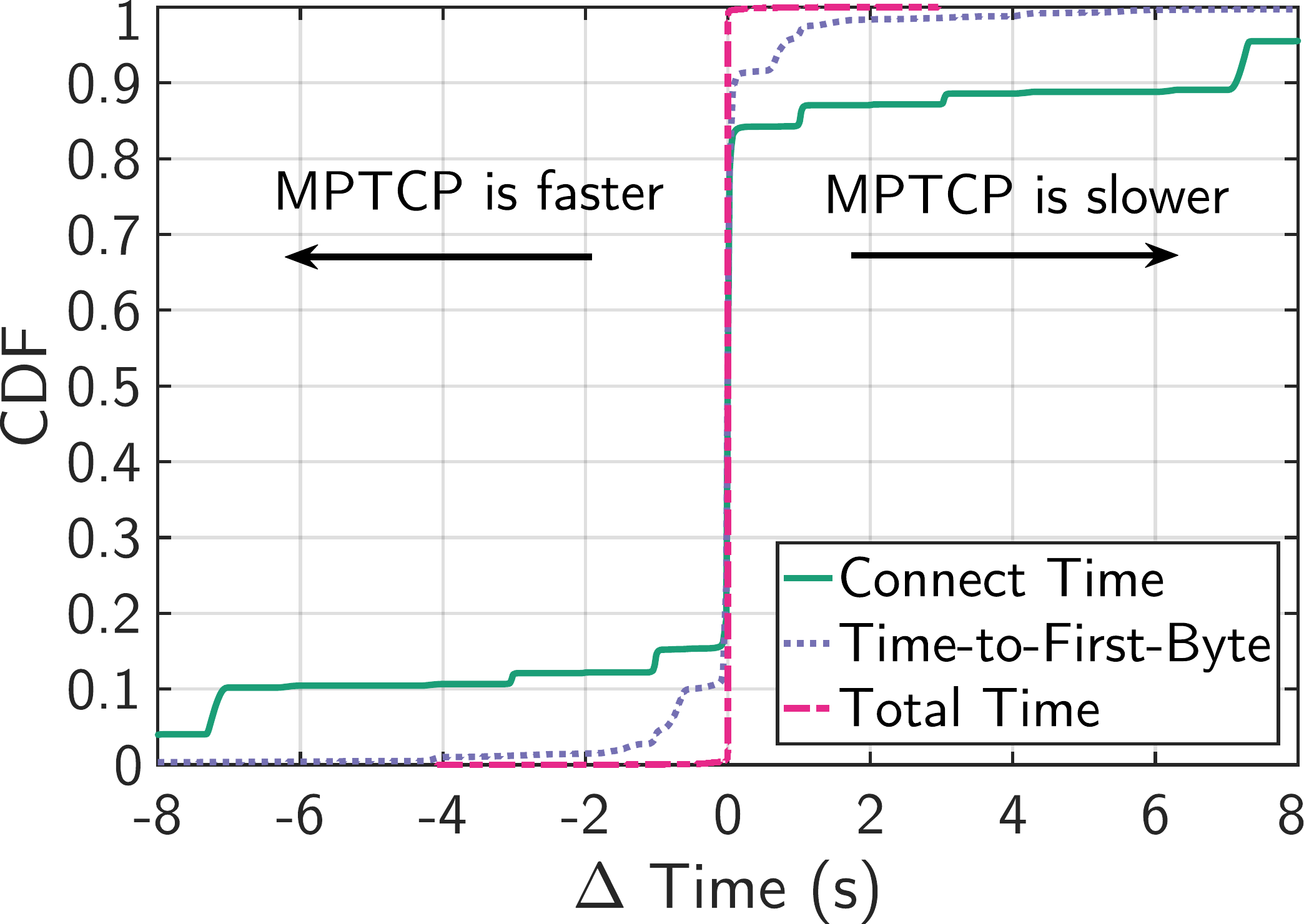}
\caption{}
    \label{fig:p80mpdiff0}
    \end{subfigure}%
    \caption{HTTP(S) GET timing deltas over MPTCP and TCP towards: \emph{truly}-MPTCP targets on port 80 (a) and port 443 (b); MPTCP targets affected by middleboxes via \tracebox (c); and targets affected by mirroring middleboxes (d). Please note the different x-axis time scale.}
    \label{fig:ipv4curlmiddlebox}
\end{figure}

\Cref{fig:p80curl,fig:p443curl} show the distribution of $\Delta$ time difference between responses from \emph{truly} MPTCP IPs identified in \Cref{sec:tracebox}.
Keep in mind that these targets are \emph{not} affected by middleboxes on the path.
$\Delta$ values less than zero denote targets that are faster using MPTCP while $\Delta > 0$ are hosts that are faster over TCP. 
Values centered around zero indicate that both protocols perform similarly.
The symmetric upper and lower distributions in \Cref{fig:p80curl} shows that the clients observe no discernible difference using (MP)TCP if connecting to targets that support MPTCP over port 80.
MPTCP-capable targets on port 443 (shown in \Cref{fig:p443curl}) show similar results for all timing values except completion time, for which the distribution tilts slightly in favor of TCP. 

We now investigate the impact of middleboxes on MPTCP traffic. 
In \Cref{fig:middleboxcurl} we show the responses from MPTCP-capable targets found to be affected by middleboxes.
As can be observed, middleboxes treat MPTCP application traffic differently.
For $\approx$30\% of \emph{all} timing values, MPTCP is slower than TCP while TCP is slower for only 10\% of measurements.
Notice the difference in x-axis ticks of \Cref{fig:middleboxcurl} and \Cref{fig:p80curl}; indicating that middleboxes can expand TTFB and load time of MPTCP connections by several seconds. 
Likely, the MPTCP client falls back to TCP before initiating data transfer for these targets since middleboxes strip away MPTCP options from the header~\cite{rfc8684}. 
As a result, such middleboxes only affect the TCP handshake phase, which also justifies large connect time values recorded for these targets.
However, not all middleboxes have a deleterious impact on MPTCP traffic, as seen in \Cref{fig:p80mpdiff0}.
The result shows that 
middleboxes that simply replay unknown TCP extensions have no discernible effect on MPTCP traffic.
Keep in mind that data transfers over these connections end up using TCP since none of the end-targets in this group were found to support MPTCP.

\begin{tcolorbox}[title=Takeaway]
  We observe no significant difference in HTTP(S) GET responses when using MPTCP over TCP from \emph{truly} MPTCP-capable servers. 
  However, we find that certain middleboxes can aggressively delay MPTCP connections, whereas TCP remains largely unaffected.
\end{tcolorbox}

\section{MPTCP Internet Traffic Share}
\label{sec:mptcpTraffic}

We quantify the real-world MPTCP traffic share by analyzing two traffic traces from geographically diverse vantage points: 
\one four years of traffic (from 2015 to 2019) on a Tier 1 ISP backbone link in North America (CAIDA traces~\cite{caida}) and
\two seven years of traffic (from 2014 to 2020) captured at the uplink of a Japanese university network (MAWI traces~\cite{mawi})

\noindent \textbf{CAIDA.} The CAIDA dataset includes bidirectional traffic captured at an Equinix data center connected to an ISP backbone link (we only consider single direction ``\textit{dir-A}" traffic in our analysis).
For 2015 and 2016, the monitor captures traffic of the ISP backbone connecting Chicago and Seattle, while for 2018 and 2019, the backbone links New York and São Paolo. 
The dataset includes a one-hour trace per month for four months of 2015 and 2016 each, ten months for 2018 and January 2019.
No data is available for 2017 and after January 2019 since the monitored links have been upgraded to 100 Gbps and exceed capturing capacity.

\noindent \textbf{MAWI.} 
The MAWI dataset includes traffic captured at samplepoint-F, a 1 Gbps transit link of the WIDE working group to an upstream ISP. 
We analyze 15 minute captures of the third Thursday of each month, from January 2014 to December 2020.
This technique allows for better comparison between months, ensuring that weekday traffic is analyzed.

Both CAIDA and MAWI datasets are anonymized, disallowing us to identify participating endpoints accurately.
However, since our objective is to understand the popularity of MPTCP in real-world Internet traffic, this does not hinder our analysis.
We remove all flows with less than five packet exchanges to prevent possible scanning traffic from influencing our study. 

\subsection{MPTCP Traffic Characteristics}

\begin{figure}[t!]
\centering
\begin{subfigure}{.49\columnwidth}
\centering
    \includegraphics[width=\columnwidth]{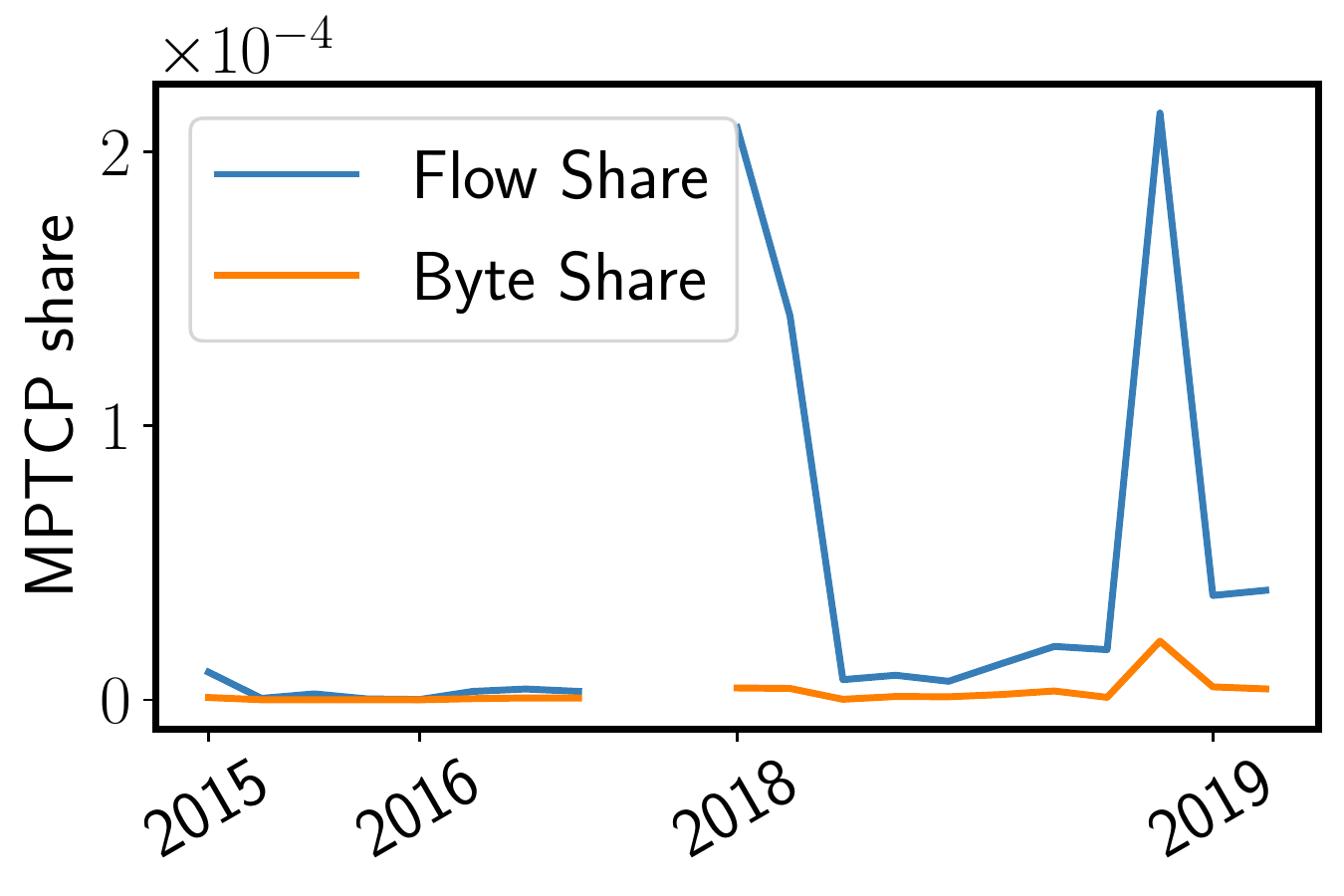}
    \caption{MPTCP traffic share.}
    \label{fig:caida-mptcp-share}
\end{subfigure}%
\begin{subfigure}{.51\columnwidth}
\centering
    \includegraphics[width=\columnwidth]{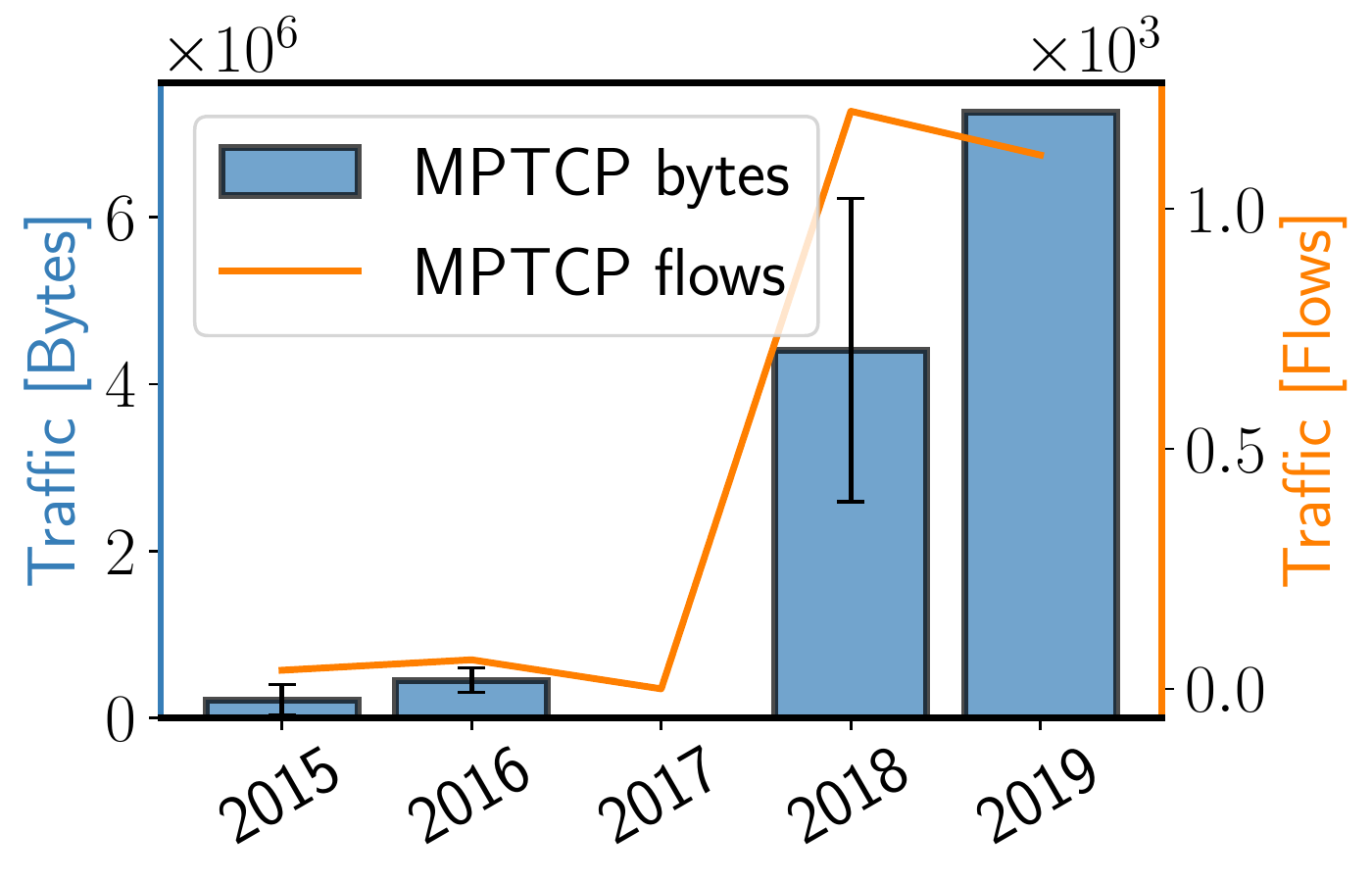}
    \caption{MPTCP absolute traffic.} 
    \label{fig:caida-mptcp-absolute}
\end{subfigure}%
    \caption{MPTCP traffic over time captured by CAIDA monitors on direction-A. (a) shows share of MPTCP flows and bytes (compared to TCP) and (b) shows their absolute values across time. The gap is due to missing data for 2017.}
    \label{fig:caida-mptcp}
\centering
\end{figure}

\begin{figure}[t!]
\begin{subfigure}{.5\columnwidth}
    \includegraphics[width=\columnwidth]{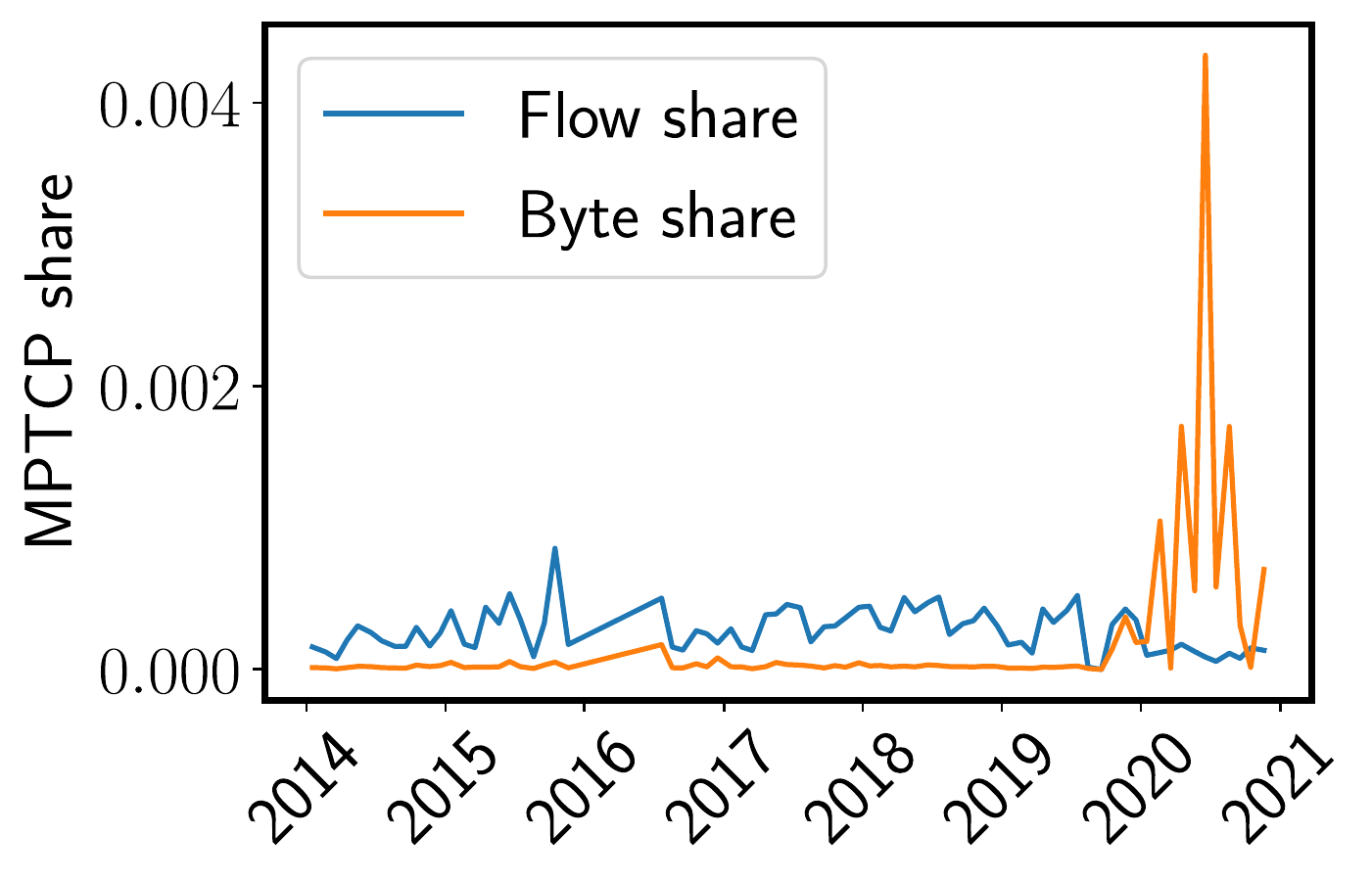}
    \caption{MPTCP traffic share.}
    \label{fig:mawi-mptcp-share}
\end{subfigure}%
\begin{subfigure}{.5\columnwidth}
    \includegraphics[width=\columnwidth]{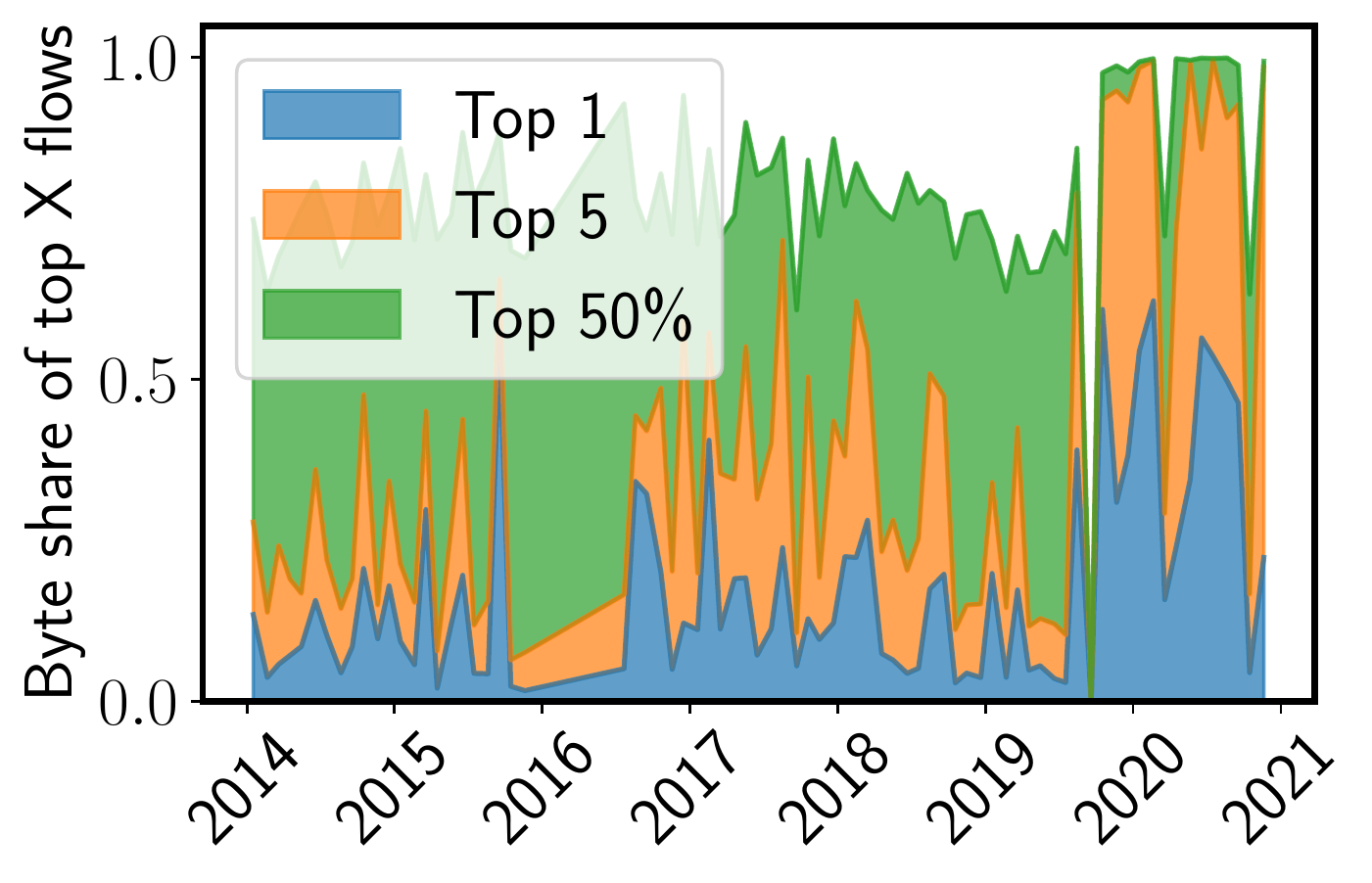}
    \caption{MPTCP flow size distribution.} 
    \label{fig:mawi-mptcp-byteshare}
    \end{subfigure}%
    \caption{MPTCP traffic over time at MAWI's samplepoint-F. (a) shows the share of MPTCP traffic (in bytes and flow) compared to TCP and (b) shows traffic share of top 1, top 5, and top 50\% flows compared to all MPTCP traffic.} 
    \label{fig:mawi-mptcp}
\end{figure}

\Cref{fig:caida-mptcp-share,fig:mawi-mptcp-share} show the share of MPTCP flows and bytes over TCP at the CAIDA and MAWI vantage points, respectively.
We observe that the MPTCP share remains fairly and consistently low in the CAIDA dataset, making up only 0.00006\% of TCP byte and 0.0003\% of TCP flow traffic.
However, there is a clear uptick in MPTCP flow share at the start of 2018 that increases as the year progresses, reaching 0.005\%.
Interestingly, the trend is mostly missing on
MPTCP byte share, indicating a simultaneous rise of TCP traffic on the link.
By the end of 2018 (and beginning of 2019), both MPTCP flow and byte share within the CAIDA dataset escalate significantly and peak at 0.02\% and 0.002\%, respectively.
\Cref{fig:caida-mptcp-absolute} paints the complementary picture of the dataset in 
absolute numbers.
The bars (attached to the left y-axis) denote the aggregate amount of MPTCP bytes, and the line (to the right y-axis) shows the mean of MPTCP flows over four years.
We observe a $\approx$~8.6$\times$ jump in MPTCP bytes from 2016--2018 and an increase of 64\% within 2018--2019.
However, the concurrent increase in the number of MPTCP flows 
hints that MPTCP is largely being used for short-lived mice transfers.
Unfortunately, we cannot analyze the after-effects of MPTCP upstreaming in Linux at the beginning of 2020 from the CAIDA dataset as no trace data is available beyond 2019.
Hence we turn our attention to the MAWI traces.

From \Cref{fig:mawi-mptcp-share} we observe that the share of MPTCP traffic flows captured by MAWI stays relatively constant over time, making up less than 0.1\% of all TCP traffic flows.
The share of MPTCP bytes is even smaller, until the end of 2019,
as we begin to see it increase significantly, peaking at upwards of 0.4\% in June 2020.
Interestingly, the number of flows remains low and does not increase. 
We further investigate this phenomenon by looking at the flow size distribution over time.
To convey this distribution, we show the traffic share of the top flow, the top five flows, and the top 50\% of flows in \Cref{fig:mawi-mptcp-byteshare}.
If all flows had the same size, the green top 50\% line would be at 0.5.
Right around the end of 2019, we see a drastic change in flow size distributions.
A single flow makes up 50\% of all MPTCP traffic at times, and the top five flows make up almost all of MPTCP traffic. 
This indicates that MPTCP is starting to be used and carries actual data.
We also evaluate the duration of these elephant flows and find that they last about 30s.
That relatively short duration also explains the few dips in the top 5 in 2020 as seen in \Cref{fig:mawi-mptcp-share}.
If an elephant MPTCP flow is not present within MAWI's 15 min capturing window, the distribution and traffic share drops.

\subsection{MPTCP Application Usage}

To better understand the applications used in MPTCP traffic, we map transport port numbers for MAWI and CAIDA traces to well-known port numbers used for specific services~\cite{ianaPorts}.
Additionally, we leverage Apple's list of ports used in their services to identify Apple service traffic \cite{applePorts}.
We are able to successfully map all flows to well-known ports in MAWI; except one flow with both high-ports and two flows with reserved value zero as the source port.
The latter could be attributed to misconfigured devices 
\cite{maghsoudlou2021zeroing}.
For CAIDA, we find more than 80\% of source ports in the well-known range and a majority of destination ports as ephemeral; indicating that the link mostly carries server-to-client upstream traffic.

Overall, we observe six different applications utilizing MPTCP in MAWI: HTTPS, HTTP, Ident, SMB, Siri, and RDP.
The overwhelming majority of all MPTCP traffic, however, is HTTPS traffic, whose lowest share is 99.5\%.
On the other hand, the application mix in the CAIDA dataset is more diverse than MAWI as we find 15+ services using MPTCP; including HTTPS, HTTP, Spamtrap, and Microsoft services.
However, similar to MAWI, HTTPS traffic eclipses all other applications with 99.91\% being its lowest share.
Moreover, other than very small traces of Siri in 2018, we did not discover any other instances of Apple services using MPTCP in both datasets.

\begin{tcolorbox}[title=Takeaway]
The MPTCP traffic share remains consistently low over time.
Since mid 2019, MPTCP traffic has shown a steady increase and now includes larger flows, indicating that MPTCP has started to see actual deployment.
With more than 99\% of all MPTCP traffic, HTTPS is the dominant application using MPTCP in-the-wild.
\end{tcolorbox}

\section{Discussion and Conclusion}\label{sec:conclusion}

This paper presented the first broad multi-faceted assessment on MPTCPv0.
We studied both the \textit{infrastructure}, by probing the entire IPv4 address space and an IPv6 hitlist for MPTCP-capable IPs, and \textit{traffic share} at two geographically diverse vantage points.
We identified middleboxes that impact both MPTCP scanning attempts and user traffic during the course of our study, hence providing the most accurate picture of \emph{true} MPTCP deployment to date.
We observed a steady growth in MPTCP-enabled IPs that support HTTP and HTTPS in our six-month investigation period, reaching $\approx$ 9k and 30+ in December 2020 for IPv4 and IPv6, respectively.
The growth is primarily driven by Apple and ISPs across the globe that rely on the protocol to enhance their services.

The rise in \emph{infrastructure} size is not yet reflected in MPTCP's \emph{traffic share} as MPTCP's byte share peaks at a low 0.4\%, however showing an apparent increase in 2020.
Combined with MPTCP's susceptibility to middleboxes, the path to wide-spread adoption of the protocol encounters several roadblocks.
We identified the presence of middleboxes, which can aggressively degrade the perceived quality of applications employing MPTCP.
Compare it to multipath alternatives being developed in parallel using substrates such as QUIC that remain largely unaffected by middleboxes, the popularity of MPTCP in the future is still unknown.
Interestingly, MPTCP seems to encounter fewer hurdles over IPv6 compared to IPv4, as the new Internet Protocol lacks middleboxes that interfere with the protocol's operation.
We envision the wide-spread adoption of MPTCP to be possible only if it is embraced by both middlebox developers and service-providing organizations simultaneously.

\noindent \textbf{Limitations:} In this study, we focus on the evaluation of MPTCPv0 deployments over the Internet.
Our methodology does not capture client-side MPTCP deployments, including MPTCP proxy solutions that work only when the client establishes an MPTCP connection. The passive data analysis (cf. \Cref{sec:mptcpTraffic}) only focuses on MPTCP support and not MPTCP usage. 
We plan to plug these limitations in a future study, with MPTCPv1 results being already available at \url{https://mptcp.io}.

\noindent \textbf{Acknowledgments:} We thank the reviewers and Olivier Bonaventure for the feedback and comments on this paper. 

\balance

\bibliographystyle{IEEEtran}
\bibliography{bibliography.bib}

\end{document}